\documentstyle[12pt,epsfig,color,rawfonts,pictex,citesort]{article}
\newcommand{\Z}{{\sf Z \!\!\! Z}}

\newcommand{\1}{{\sf 1 \!\! 1}}

\newcommand{\p}{\partial}

\makeatletter
\@addtoreset{equation}{section}
\makeatother

\title{The Deconfinement Phase Transition of \\
$Sp(2)$ and $Sp(3)$ Yang-Mills Theories \\
in $2+1$ and $3+1$ Dimensions}

\author{K. Holland$^{{\rm a}}$, M. Pepe$^{{\rm b}}$,
and U.-J. Wiese$^{{\rm b}}$\footnote{on leave from MIT}
\\ \\
$^{{\rm a}}$ Department of Physics, University of California at San Diego \\
La Jolla, CA 92093, U.S.A. \\ \\
$^{{\rm b}}$ Institute for Theoretical Physics \\
Bern University, Sidlerstrasse 5, CH-3012 Bern, Switzerland \\}

\begin{document}

\maketitle

\vspace{-1cm}

\begin{abstract} \normalsize

Some time ago, Svetitsky and Yaffe have argued that --- if the deconfinement 
phase transition of a $(d+1)$-dimensional Yang-Mills theory with gauge group
$G$ is second order --- it should be in the universality class of a 
$d$-dimensional spin model symmetric under the center of $G$. For $d = 3$ these
arguments have been confirmed numerically only in the $SU(2)$ case with center
$\Z(2)$, simply because all $SU(N)$ Yang-Mills theories with $N \geq 3$ seem to
have non-universal first order phase transitions. The symplectic groups $Sp(N)$
also have the center $\Z(2)$ and provide another extension of $SU(2) = Sp(1)$ 
to general $N$. Using lattice simulations, we find that the deconfinement phase
transition of $Sp(2)$ Yang-Mills theory is first order in $3+1$ dimensions,
while in $2+1$ dimensions stronger fluctuations induce a second order 
transition. In agreement with the Svetitsky-Yaffe conjecture, for $(2+1)$-d 
$Sp(2)$ Yang-Mills theory we find the universal critical behavior of the 2-d 
Ising model. For $Sp(3)$ Yang-Mills theory the transition is first order both 
in $2+1$ and in $3+1$ dimensions. This suggests that the size of the gauge
group --- and not the center symmetry --- determines the order of the
deconfinement phase transition.

\end{abstract}
\newpage 

\section{Introduction}

The $SU(N)$ Yang-Mills theory has a $\Z(N)$ center symmetry \cite{tHo77,tHo79}
that is spontaneously broken at high temperatures. The corresponding order 
parameter is the Polyakov loop \cite{Pol78,Sus79} whose expectation value 
$\langle \Phi \rangle = \exp(- \beta F)$ determines the free energy $F$ of a 
static quark as a function of the inverse temperature $\beta = 1/T$. In the 
low-temperature confined phase the center symmetry is unbroken, i.e.\ 
$\langle \Phi \rangle = 0$, and hence the free energy of a single static quark 
is infinite. In the high-temperature deconfined phase, on the other hand, the 
center symmetry is spontaneously broken, i.e.\ $\langle \Phi \rangle \neq 0$, 
and the free energy of a quark is finite. If the deconfinement phase transition
is second order, its long-range physics is dominated by fluctuations in the 
Polyakov loop order parameter. In this case the details of the underlying
dynamics become irrelevant and only the center symmetry and the dimensionality 
of space determine the universality class. As was first pointed out by 
Svetitsky and Yaffe, for an $SU(N)$ Yang-Mills theory in $d+1$
space-time dimensions the effective theory describing the fluctuations
of the order parameter is a  $d$-dimensional $\Z(N)$-symmetric scalar
field theory for the Polyakov loop \cite{Sve82}. For recent reviews on
this subject we refer the reader to refs.\ \cite{Hol01,Gre03}.
 
For $(d+1)$-dimensional $SU(2)$ Yang-Mills theory the center symmetry is 
$\Z(2)$ and hence the effective theory is a $d$-dimensional $\Z(2)$-symmetric 
scalar field theory for the real-valued Polyakov loop. The simplest theory in 
this class is a $\Phi^4$ theory with the Euclidean action
\begin{equation}
S[\Phi] = \int d^dx \ \left[\frac{1}{2} \p_i \Phi \p_i \Phi + V(\Phi)\right].
\end{equation}
The scalar potential is given by
\begin{equation}
V(\Phi) = a \Phi^2 + b \Phi^4,
\end{equation}
where $b > 0$ for stability reasons. Indeed, for $a = 0$ this theory has a 
second order phase transition in the universality class of the $d$-dimensional 
Ising model. However, this does not guarantee that the deconfinement phase 
transition in $SU(2)$ Yang-Mills theory is also second order. In particular, 
one could imagine that the effective potential 
\begin{equation}
\label{effpot}
V(\Phi) = a \Phi^2 + b \Phi^4 + c \Phi^6,
\end{equation}
also involves a $\Phi^6$ term which is marginally relevant in three dimensions.
Then the coefficient $c$ has to be positive in order to ensure that the 
potential is bounded from below, but the coefficient $b$ of the $\Phi^4$ term 
can now become negative. Then the phase transition becomes first order. For 
$a = b = 0$ there is a tricritical point at which the order of the phase
transition changes. Still, this does not happen in $(3+1)$-d $SU(2)$ Yang-Mills
theory, where --- using lattice simulations --- it has indeed been shown that 
the deconfinement phase transition is second order 
\cite{McL81a,McL81b,Kut81,Eng81,Gav83a,Gav83b} and has the same critical 
exponents as the 3-d Ising model \cite{Eng90,Eng92}. Similarly, $(2+1)$-d
$SU(2)$ Yang-Mills theory has a second order deconfinement phase transition
\cite{Tep93} in the universality class of the 2-d Ising model \cite{Eng97}.

For $N \geq 3$, the effective theory for a $(d+1)$-dimensional 
$SU(N)$ Yang-Mills theory with the center $\Z(N)$ should be a $d$-dimensional 
$\Z(N)$-symmetric scalar field theory for the complex-valued Polyakov loop 
$\Phi = \Phi_1 + i \Phi_2$ \cite{Yaf82}. A simple representative of this class 
of theories is defined by the action
\begin{equation}
S[\Phi] = \int d^dx \ \left[\frac{1}{2} \p_i \Phi^* \p_i \Phi + V(\Phi)\right],
\end{equation}
with
\begin{equation}
V(\Phi) = a |\Phi|^2 + b |\Phi|^4 + c |\Phi|^6 + d \ \mbox{Re}(\Phi^N).
\end{equation}
Note that $\mbox{Im}(\Phi^N)$ is also $\Z(N)$ invariant, but not invariant 
under charge conjugation. Hence, this term cannot appear in the effective 
action. 

For $N = 3$ the cubic term in the action
\begin{equation}
\mbox{Re}(\Phi^3) = \Phi_1(\Phi_1^2 - 3 \Phi_2^2),
\end{equation}
breaks the $U(1)$ symmetry of the quadratic and quartic terms down to $\Z(3)$. 
For $d = 3$, the presence of this term renders the phase transition first order
\cite{Yaf82}. Also the 3-d 3-state Potts model 
\cite{Kna79,Blo79,Her79,Wu82,Fuk89b,Gav89} has a first order phase transition. The
absence of universal behavior in 3-d $\Z(3)$-symmetric models suggests that the
deconfinement phase transition in $SU(3)$ Yang-Mills theory should also be 
first order. Indeed this has been confirmed in great detail in lattice
simulations \cite{Cel83,Kog83,Got85,Bro88,Fuk89a,Alv90}. In $(2+1)$-d $SU(3)$
Yang-Mills theory stronger fluctuations lead to a second order phase transition
in the universality class of the 2-d 3-state Potts model \cite{Chr92}.

Next, we consider the $N = 4$ case. For $c > 0$ the $\Z(4)$-symmetric scalar 
potential leads to a second order phase transition in the universality class of
the 3-d $\Z(4)$-symmetric chiral clock model which corresponds to two decoupled
Ising models. However, the deconfinement phase transition of $(3+1)$-d $SU(4)$ 
Yang-Mills theory does not seem to fall into that universality class. It is 
inconvenient to study the deconfinement phase transition in lattice simulations
of $SU(4)$ Yang-Mills theory due to a first order bulk phase transition at zero
temperature. The existing lattice data show that the deconfinement transition 
is first order \cite{Whe84,Goc84,Bat84,Win01,Tep02}. Again, in
$(2+1)$-d $SU(4)$ Yang-Mills theory stronger fluctuations seem to
induce a second order phase transition \cite{Gro85,deF03b}.

For $N \geq 5$, depending on the values of the various parameters in the
potential $V(\Phi)$, the phase transition can again be first or second order. 
In case of a second order phase transition, the corresponding scalar field 
theory should be in the universality class of the $\Z(N)$-symmetric chiral 
clock model which, for $d = 3$, happens to be the one of the $U(1)$-symmetric 
XY-model \cite{Hov03}. Thus, for $N \geq 5$, the discrete $\Z(N)$ symmetry is 
not visible at the critical point and is, in fact, dynamically enhanced to a 
continuous $U(1)$ symmetry. This should not be too surprising. In particular, 
for $N \geq 7$ the term $\mbox{Re}(\Phi^N)$ which breaks the $U(1)$ symmetry 
of the other terms down to $\Z(N)$ is irrelevant in three dimensions. However,
again numerical simulations --- in this case of $SU(6)$ and 
$SU(8)$ Yang-Mills theory \cite{Tep02} --- indicate a first order deconfinement
phase transition. This suggests that all $(3+1)$-d $SU(N)$ Yang-Mills theories 
with $N \geq 3$ have a first order deconfinement phase transition without 
universal behavior. In particular, the universality arguments of \cite{Sve82} 
then apply only to the $SU(2)$ case, not to $N \geq 3$ or to the $N = \infty$ 
limit. It should be mentioned that other arguments may suggest a second order 
phase transition at large $N$ \cite{Pis97}.

It is interesting to ask if Svetitsky and Yaffe's universality arguments can be
applied beyond $SU(2)$ Yang-Mills theory, along another direction in the space 
of Lie groups. In particular, since $SU(2) \simeq SO(3)$, one can ask if 
$SO(N)$ Yang-Mills theories show universal behavior at their deconfinement 
phase transitions. Numerical studies of $(3+1)$-d $SO(3)$ gauge theory are 
complicated by a bulk phase transition in which the lattice theory sheds off 
its $\Z(2)$ center monopole lattice artifacts 
\cite{Gre81,Bha81,Hal81a,Hal81b,Can82}. Beyond this phase transition, in the 
continuum limit, one would expect $SO(3)$ and $SU(2)$ Yang-Mills theories to be
equivalent. Lattice studies of the deconfinement phase transition of $SO(3)$ 
Yang-Mills theory \cite{Che96,Dat98,deF03,Bar03} are consistent with this 
expectation.

In order to avoid complications due to lattice artifacts, it is best to work
with the universal covering group of $SO(N)$ which is $Spin(N)$. For example,
$Spin(3) = SU(2)$. The center of $Spin(N)$ is $\Z(2)$ for odd $N$, $\Z(2) 
\otimes \Z(2)$ for $N = 4k$, and $\Z(4)$ for $N = 4k + 2$. Let us first discuss
the family of $Spin(N)$ with odd $N$ and center $\Z(2)$. The simplest case is
$Spin(3) = SU(2)$ which we already discussed. Since $Spin(5) = Sp(2)$, this is 
a case that we will concentrate on later in this paper. In contrast to 
$Spin(3)$, for $d = 3$, we find a first order deconfinement phase transition.
We are unaware of numerical lattice studies of $Spin(N)$ Yang-Mills theories 
with $N \geq 7$. Next, we consider $Spin(N)$ with $N = 4k$ and center 
$\Z(2) \otimes \Z(2)$. Now the simplest case is $Spin(4) = SU(2) \otimes 
SU(2)$. A $Spin(4)$ lattice Yang-Mills theory with the standard Wilson action 
factorizes into two $SU(2)$ Yang-Mills theories. Hence, its deconfinement phase
transition is in the universality class of two decoupled Ising models. The next
case in this family is $Spin(8)$ which has not been studied numerically. The 
last family is $Spin(N)$ with $N = 4k + 2$ with the center $\Z(4)$. Then the 
simplest case is $Spin(6) = SU(4)$. As we already discussed, lattice 
simulations have shown that in $(3+1)$-d $SU(4)$ Yang-Mills theory the 
deconfinement phase transition is first order. Although this need not 
necessarily be the case for larger $N$, due to the increasing number of gauge 
degrees of freedom we expect $(3+1)$-d $Spin(N)$ gauge theories to have first 
order transitions for all $N \geq 5$.

There is a last possible direction in the space of Lie groups which we explore
in this paper. This is the sequence of symplectic Lie groups $Sp(N)$ which are
simply connected and hence are their own universal covering groups. This 
sequence has the center $\Z(2)$ for all $N$ and includes $SU(2) = Sp(1)$. The
groups $Sp(N)$ are pseudo-real and thus provide a natural extension of $SU(2)$
to larger $N$. In contrast to $SU(N)$, the study of $Sp(N)$ Yang-Mills theories
allows us to change the size of the group without changing the center. Hence,
we can investigate the order of the deconfinement phase transition as a 
function of the size of the gauge group, keeping the available Ising 
universality class fixed. In fact, we will argue that the order of the phase 
transition is controlled by the size of the gauge group and not by the center
symmetry. Our studies of $(3+1)$-d $Sp(2)$ and $Sp(3)$ Yang-Mills theories show
that a first order phase transition arises although the 3-d Ising universality 
class is available. The order of the phase transition is 
a dynamical issue which does not simply follow from the center symmetry. The 
Lie groups larger than $SU(2)$ have many generators and thus give rise to a 
large number of deconfined gluons. The number of confined glueball states, on 
the other hand, is essentially independent of the gauge group. Hence, there is 
a drastic change in the number of relevant degrees of freedom on the two sides 
of the deconfinement phase transition. This can induce the abrupt changes in 
thermodynamical quantities that are characteristic for a first order phase 
transition. This suggests that the deconfinement phase transition of $(3+1)$-d 
$Sp(N)$ Yang-Mills theory is first order for all $N \geq 2$. In $2+1$ 
dimensions we find that stronger fluctuations drive the phase transition of 
$Sp(2)$ Yang-Mills theory second order. Due to the larger number of gauge 
degrees of freedom, $(2+1)$-d $Sp(3)$ Yang-Mills theory still has a first order
phase transition. We expect this to be the case for all $N \geq 3$.

For completeness, let us also discuss the exceptional Lie groups $G(2)$, 
$F(4)$, $E(6)$, $E(7)$, and $E(8)$ that do not fall in the main sequences 
$SU(N)$, $Spin(N)$, or $Sp(N)$. The groups $G(2)$, $F(4)$, and $E(8)$ have a 
trivial center and thus need not have a deconfinement phase transition at all.
In fact, recently we have argued that $G(2)$ Yang-Mills theory should only have
a crossover between its low- and high-temperature regimes \cite{Hol03}. 
The group $E(6)$ has the center $\Z(3)$. Just as in the $SU(3)$ case,
a $(3+1)$-d $E(6)$ Yang-Mills theory is expected to have a first order
deconfinement phase transition because no universality class with
$\Z(3)$ symmetry seems to exist in three dimensions. However, even for
$(2+1)$-d $E(6)$ Yang-Mills theory, where the 2-d 3-state Potts model
universality class is available, we expect a first order phase
transition due to the large size of $E(6)$ (which has rank 6 and 78
generators). Finally, $E(7)$ has the center $\Z(2)$. If the
deconfinement phase transition of $E(7)$ Yang-Mills theory is second
order, it should hence have Ising critical exponents. However, for a
group as large as $E(7)$ (with rank 7 and 133 generators) the large
number of gauge degrees of freedom again suggests a first order
deconfinement phase transition.  

Also taking into account the numerical results for the various small Lie 
groups, the arguments from above lead us to conjecture that, in $3+1$ 
dimensions, only $SU(2)$ and its trivial extension
$Spin(4)=SU(2)\otimes SU(2)$ Yang-Mills theories have a second order
deconfinement phase transition. All other Yang-Mills theories are
expected to have a first order phase transition; possible exceptions
are Yang-Mills theories with gauge group $G(2)$, $F(4)$, and $E(8)$
which have a trivial center and may hence just have a crossover. In
that case, in $3+1$ dimensions Svetitsky and Yaffe's universality
arguments apply only to $SU(2)$.   

In $2+1$ dimensions stronger fluctuations arise and may result in a second 
order phase transition. Indeed, for $(2+1)$-d $Sp(2)$ Yang-Mills theory we find
a second order phase transition. In agreement with Svetitsky and Yaffe's 
arguments, a finite-size scaling analysis shows critical behavior in the 2-d 
Ising universality class. For the first time, this confirms the Svetitsky-Yaffe
conjecture beyond $SU(N)$. For $(2+1)$-d $Sp(3)$ Yang-Mills theory, on the 
other hand, the transition is first order. This is again consistent with the 
increasing number of gauge degrees of freedom. Hence, in $2+1$ dimensions we 
expect the transition to be first order for all $N \geq 3$. 

The rest of this paper is organized as follows. In section 2 basic properties 
of $Sp(N)$ groups are reviewed. Section 3 introduces $Sp(N)$ Yang-Mills theory 
on the lattice, including heatbath and overrelaxation algorithms for its
numerical simulation. Evidence for a first order deconfinement phase transition
is reported for both $Sp(2)$ and $Sp(3)$ Yang-Mills theory in $3+1$ dimensions 
and for $Sp(3)$ in $2+1$ dimensions. A finite-size scaling analysis of the 
deconfinement phase transition in $(2+1)$-d $Sp(2)$ Yang-Mills theory shows 
that it is second order and in the universality class of the 2-d Ising model. 
We also study the static quark potential in $(3+1)$-d $Sp(2)$ Yang-Mills theory
using the multi-level algorithm introduced by L\"uscher and Weisz \cite{Lue01}.
The resulting string tension is used to express the critical temperature of the
deconfinement phase transition in physical units. Finally, section 4 contains 
our conclusions. Summaries of this work have already appeared in \cite{Hol03a,Hol03b}.

\section{The Symplectic Group $Sp(N)$}

The group $Sp(N)$ is a subgroup of $SU(2N)$ which leaves the skew-symmetric 
matrix
\begin{equation}
J = \left(\begin{array}{cc} 0 & \1 \\ - \1 & 0 \end{array}\right) = 
i \sigma_2 \otimes \1,
\end{equation}
invariant. Here $\1$ is the $N \times N$ unit-matrix and $\sigma_2$ is the
imaginary Pauli matrix. The elements $U \in SU(2N)$ that belong to the subgroup
$Sp(N)$ satisfy the constraint
\begin{equation}
\label{pseudoreal}
U^* = J U J^\dagger.
\end{equation}
Consequently, $U$ and $U^*$ are related by the unitary transformation $J$. 
Hence the $2N$-dimensional fundamental representation of $Sp(N)$ is 
pseudo-real. The matrix $J$ itself also belongs to $Sp(N)$. This implies that 
in $Sp(N)$ Yang-Mills theory charge conjugation is just a global gauge 
transformation. This property is familiar from $SU(2) = Sp(1)$ Yang-Mills 
theory.

Indeed, matrices that obey the constraint eq.(\ref{pseudoreal}) form a group 
because for $U, V \in Sp(N)$ we have
\begin{equation}
(U V)^* = U^* V^* = J U J^\dagger J V J^\dagger = J (U V) J^\dagger.
\end{equation}
The inverse $U^\dagger$ also obeys the constraint because
\begin{equation}
(U^\dagger)^* = (U^*)^\dagger = (J U J^\dagger)^\dagger = 
J U^\dagger J^\dagger,
\end{equation}
and obviously the unit-matrix also belongs to $Sp(N)$.
The constraint eq.(\ref{pseudoreal}) implies the following form of a generic
$Sp(N)$ matrix
\begin{equation}
\label{group}
U =  \left(\begin{array}{cc} W & X \\ - X^* & W^* \end{array}\right), 
\end{equation}
where $W$ and $X$ are complex $N \times N$ matrices. Since $U$ must still
be an element of $SU(2N)$, these matrices must satisfy $W W^\dagger + 
X X^\dagger = \1$ and $W X^T = X W^T$. Note that the eigenvalues of $U$ come in
complex conjugate pairs. Since center elements are multiples of the 
unit-matrix, in this case eq.(\ref{group}) immediately implies $W = W^*$. 
Hence, the center of $Sp(N)$ is $\Z(2)$.

Writing $U = \exp(i H)$, where $H$ is a Hermitean traceless matrix,
eq.(\ref{pseudoreal}) implies that the generators $H$ of $Sp(N)$ satisfy the
constraint
\begin{equation}
\label{constraint}
H^* = - J H J^\dagger = J H J.
\end{equation}
This relation leads to the following generic form,
\begin{equation}
H =  \left(\begin{array}{cc} A & B \\ B^* & - A^* \end{array}\right),
\end{equation}
where $A$ and $B$ are $N \times N$ matrices. The Hermiticity condition $H =
H^\dagger$ implies $A = A^\dagger$ and $B = B^T$. Note that, since $A$ is
Hermitean, $H$ is automatically traceless. The Hermitean $N \times N$ matrix 
$A$ has $N^2$ degrees of freedom and the complex symmetric $N \times N$ matrix
$B$ has $(N + 1)N$ degrees of freedom. Hence the dimension of the group $Sp(N)$
is $N^2 + (N + 1)N = (2 N + 1)N$. There are $N$ independent diagonal generators
of the maximal Abelian Cartan subgroup. Hence the rank of $Sp(N)$ is $N$. The 
$N = 1$ case is equivalent to $SU(2)$, while the $N = 2$ case is equivalent to 
$SO(5)$, or more precisely to its universal covering group $Spin(5)$. Since 
$Sp(2)$ has rank 2, the weight diagrams of its representations can be drawn in 
a 2-d plane. The weight diagrams of the fundamental representation $\{4\}$, the
$SO(5)$ vector representation $\{5\}$, and the adjoint representation $\{10\}$ 
are depicted in figures 1, 2, and 3, respectively.

\begin{figure}[htb]
\begin{center}
\epsfig{figure=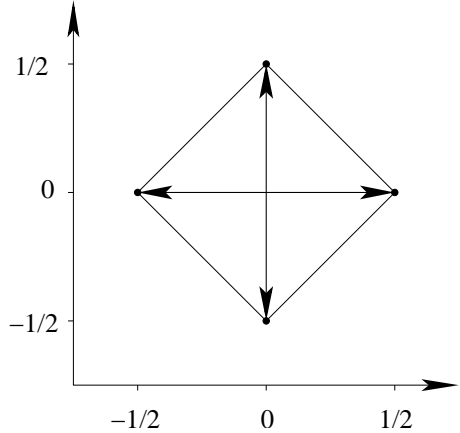,width=6cm}
\end{center}
\caption{\it The weight diagram for the fundamental $\{4\}$ representation of
$Sp(2)$.}
\end{figure}

\begin{figure}[htb]
\begin{center}
\epsfig{figure=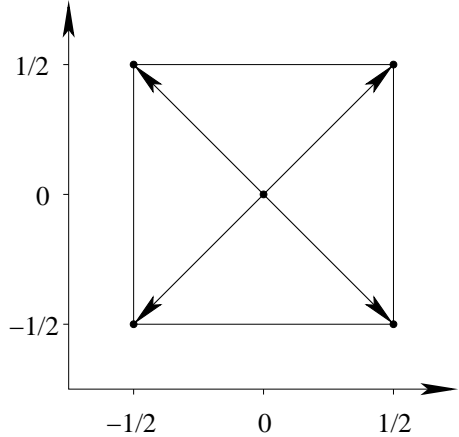,width=6cm}
\end{center}
\caption{\it The weight diagram for the $\{5\}$ representation of
$Sp(2)$ (the vector representation of $SO(5)$).}
\end{figure}

\begin{figure}[htb]
\begin{center}
\epsfig{figure=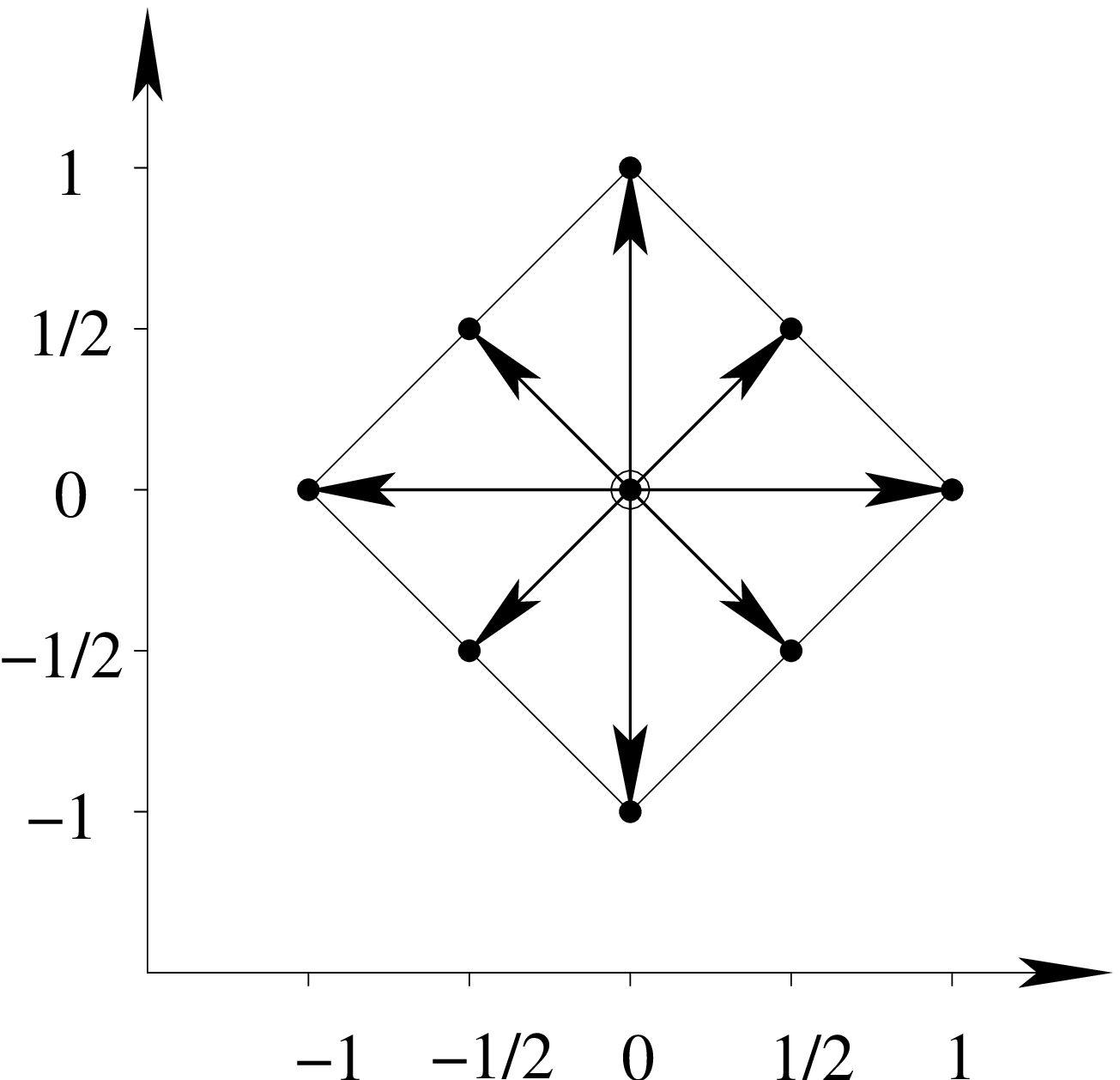,width=6cm}
\end{center}
\caption{\it The weight diagram for the adjoint $\{10\}$ representation of
$Sp(2)$.}
\end{figure}

\section{$Sp(N)$ Yang-Mills Theory on the Lattice}

In this section we consider $Sp(N)$ Yang-Mills theory on the lattice. First, we
discuss the action, the measure, and important observables. Then we describe
the simulation techniques and present results of numerical computations.

\subsection{Action, Measure, and Observables}

The construction of $Sp(N)$ Yang-Mills theory on the lattice is 
straightforward. The link parallel transporter matrices $U_{x,\mu} \in Sp(N)$ 
are group elements in the fundamental $\{2N\}$ representation. We consider the 
standard Wilson plaquette action
\begin{equation}
S[U] = - \frac{2}{g^2} \sum_\Box \mbox{Tr} \ U_\Box =
- \frac{2}{g^2} \sum_{x,\mu <\nu} \mbox{Tr} \
(U_{x,\mu} U_{x+\hat\mu,\nu} U^\dagger_{x+\hat\nu,\mu} U^\dagger_{x,\nu}),
\end{equation}
where $g$ is the bare gauge coupling. The partition function then takes the 
form
\begin{equation}
Z = \int {\cal D}U \exp(- S[U]),
\end{equation}
where the path integral measure
\begin{equation}
\int {\cal D}U = \prod_{x,\mu} \int_{Sp(N)} dU_{x,\mu},
\end{equation}
is a product of local Haar measures of the group $Sp(N)$ for each link. Both 
the action and the measure are invariant under gauge transformations
\begin{equation}
U'_{x,\mu} = \Omega_x U_{x,\mu} \Omega^\dagger_{x+\hat\mu},
\end{equation}
with $\Omega_x \in Sp(N)$. The Polyakov loop
\begin{equation}
\Phi_{\vec x} = \mbox{Tr}({\cal P} \prod_{t = 1}^{N_t} U_{\vec x,t,d+1})
\end{equation}
is the trace of a path ordered product of link variables along a loop wrapping 
around the periodic Euclidean time direction. Here $N_t = 1/T$ is the extent of
the lattice in Euclidean time, which determines the temperature $T$ in lattice
units. The lattice action is invariant under global $\Z(2)$ center symmetry 
transformations
\begin{equation}
U_{\vec x,N_t,d+1}' = - U_{\vec x,N_t,d+1},
\end{equation}
while the Polyakov loop changes sign, i.e.\ $\Phi_{\vec x}' = - \Phi_{\vec x}$.
The expectation value of the Polyakov loop is given by
\begin{equation}
\langle \Phi \rangle = 
\frac{1}{Z} \int {\cal D}U \ \frac{1}{L^d} \sum_{\vec x} \Phi_{\vec x} 
\exp(- S[U]).
\end{equation}
Here $L^d$ is the spatial lattice volume, again in lattice units. As a 
consequence of the $\Z(2)$ center symmetry, with periodic boundary conditions 
in the spatial directions the expectation value of the Polyakov loop
always vanishes, i.e.\ $\langle \Phi \rangle = 0$, even in the
deconfined phase. This is simply because spontaneous symmetry breaking
--- in the sense of a non-vanishing order parameter --- does not occur
in a finite volume. Alternatively, one may say  that the expectation
value of the Polyakov loop vanishes because the presence of a single
static quark is incompatible with the Gauss law on a torus
\cite{Hil83}. Since it always vanishes, on a finite torus the
expectation value of the Polyakov loop does not contain any useful
information about confinement or deconfinement. In the finite-size
scaling analysis presented below we therefore consider the expectation
value of the magnitude of the Polyakov loop $\langle |\Phi| \rangle$. 
This quantity is always non-zero in a finite volume, but vanishes in
the confined phase in the infinite volume limit. Furthermore, using
the Polyakov  loop, one can define its probability distribution  
\begin{equation}
p(\Phi) = \frac{1}{Z} \int {\cal D}U \ \delta\left(\Phi - \frac{1}{L^d} 
\sum_{\vec x} \mbox{Tr}({\cal P} \prod_{t = 1}^{N_t} U_{\vec x,t,d+1})\right)
\exp(- S[U]),
\end{equation}
which does indeed allow one to distinguish confined from deconfined phases. In
the confined phase $p(\Phi)$ has a single maximum at $\Phi = 0$, while in the
deconfined phase it has two degenerate maxima at $\Phi \neq 0$. If the
deconfinement phase transition is first order, the confined and the two 
deconfined phases coexist and one can simultaneously observe three maxima 
close to the phase transition. At a second order phase transition, on the other
hand, the high- and low-temperature phases become indistinguishable. The two 
maxima of the deconfined phase then merge and smoothly turn into the single 
maximum of the confined phase. Three coexisting maxima then do not occur in a 
large volume. 

In a pure glue theory another quantity of physical interest is the static
quark potential $V_{QQ}(\vec R)$. Note that, since the fundamental 
representation of $Sp(N)$ is pseudo-real, in $Sp(N)$ gauge theories quarks and
anti-quarks are indistinguishable. At any temperature, $V_{QQ}(\vec R)$ can 
be derived from the 2-point correlation function of the Polyakov loop
\begin{equation}
\langle \Phi_{\vec x} \Phi_{\vec y} \rangle = 
\exp(- N_t V_{QQ}(\vec x - \vec y)).
\end{equation}
In the zero-temperature limit, $N_t \rightarrow \infty$, at large distances
$R = |\vec R| = |\vec x - \vec y|$ the static quark potential is linearly 
rising, 
\begin{equation}
\label{pot}
V_{QQ}(\vec R) \sim V_0 + \frac{c}{R} + \sigma R,
\end{equation}
with the slope given by the string tension $\sigma$. At finite temperature 
$V(\vec R)$ measures the free energy of a static quark pair at the distance 
vector $\vec R$. The string tension is a temperature-dependent function which 
decreases with increasing temperature. In the deconfined phase it vanishes and 
the potential levels off.

As physical quantities that are useful in the finite-size scaling analysis 
presented below we also introduce the Polyakov loop susceptibility
\begin{equation}
\chi = \sum_{\vec x} \langle \Phi_{\vec 0} \Phi_{\vec x} \rangle
= L^d \langle \Phi^2 \rangle,
\end{equation}
as well as the Binder cumulant \cite{Bin81}
\begin{equation}
g_R = \frac{\langle \Phi^4 \rangle}{\langle \Phi^2 \rangle^2} - 3.
\end{equation}
The susceptibility measures the strength of fluctuations in the order parameter
while the Binder cumulant measures the deviation from a Gaussian distribution
of those fluctuations. We also consider the specific heat which takes the form
\begin{equation}\label{spec-heat}
C_V = 
\frac{1}{L^d N_t} (\langle S^2 \rangle - \langle S \rangle^2).
\end{equation}
In a finite volume $C_V$ has a maximum close to the critical coupling
$4N/g_c^2$ of the infinite 
volume theory. We denote the value of the specific heat at the maximum
by $C_V^{max}$. Another interesting observable is the latent heat 
\begin{equation}\label{lat-heat}
L_H = \frac{1}{L^d N_t}(\langle S \rangle_c - \langle S \rangle_d),
\end{equation}
which measures the difference of the expectation values of the action
in the confined and the deconfined phase. Note that this quantity is
defined only at the phase transition in the infinite volume limit. It
vanishes for a second order phase transition and is non-zero for a
first order transition. Even in a finite (but sufficiently large)
volume, $L_H$ can be evaluated in a Monte Carlo simulation because ---
up to very rare tunneling events --- every configuration can be
unambiguously associated with the confined or the deconfined phase. In
the large volume limit, the latent heat and the maximum of the
specific heat are related by \cite{Bil92}  
\begin{equation}\label{speclat-heat}
C_V^{max} = L ^d N_t \frac{L_H^2 }{4}.
\end{equation}

\subsection{Simulation Techniques and Basic Results}

The $Sp(N)$ lattice Yang-Mills theory with the standard Wilson action can be
simulated with heat-bath \cite{Cre80} and microcanonical overrelaxation 
\cite{Adl81,Cre87,Bro87} algorithms similar to the ones for
$SU(N)$. The main idea, originally due to Cabibbo and Marinari \cite{Cab82}, is
to work sequentially in various $SU(2) = Sp(1)$ subgroups. In the $Sp(2)$ case
we use four different $SU(2)$ subgroups: two of them operating along the two
axes in the weight diagram in figure 1, and two of them operating along the two
diagonals. Under the first two $SU(2)$ subgroups the four states of the
fundamental $Sp(2)$ representation decompose into one $SU(2)$ doublet and two 
$SU(2)$ singlets $\{4\} = \{2\}_{SU(2)} \oplus \{1\}_{SU(2)} \oplus 
\{1\}_{SU(2)}$. Under the other two $SU(2)$ subgroups the fundamental 
representation decomposes into two $SU(2)$ doublets $\{4\} = \{2\}_{SU(2)} 
\oplus \{2\}_{SU(2)}$. To ensure ergodicity, two $SU(2)$ subgroups (one from 
each pair) are sufficient. For the general $Sp(N)$ case a minimal set of $N$ 
appropriately chosen $SU(2)$ subgroups is sufficient.

We have implemented the heat-bath and microcanonical overrelaxation algorithms
for $Sp(2)$ and $Sp(3)$. First, we have looked for a potential bulk phase 
transition at zero temperature separating a strong from a weak coupling
confined phase. Fortunately, in contrast to $SU(N)$ with $N \geq 4$, both in
$Sp(2)$ and in $Sp(3)$ no bulk phase transition (which could obscure the finite
temperature transition) has been found. In particular, different hot and cold 
starts did not show metastability. The absence of a bulk phase transition makes
it easier to take the continuum limit than, for example, in simulations of 
$SU(4)$ Yang-Mills theory. Our Monte Carlo data for the expectation value of 
the plaquette are compared with analytic weak and strong coupling expansions in
figures 4 and 5. At leading order of the weak coupling expansion, for
the plaquette expectation value one finds
\begin{equation}
\frac{1}{2N} \langle \mbox{Tr} U_\Box \rangle 
= 1 - \frac{(2N + 1) g^2}{4(d + 1)}
+{\cal{O}} ( g^4 ) ,
\end{equation}
(where $d+1$ is the dimension of space-time), while at strong coupling
\begin{equation}
\frac{1}{2N} \langle \mbox{Tr} U_\Box \rangle = \frac{1}{N g^2} +
{\cal{O}} \left( \frac{1}{g^{10}} \right) .
\end{equation}
For comparison with potential future studies, some of our Monte Carlo data are 
listed in table 1.

\begin{figure}[htb]
\begin{center}
\epsfig{figure=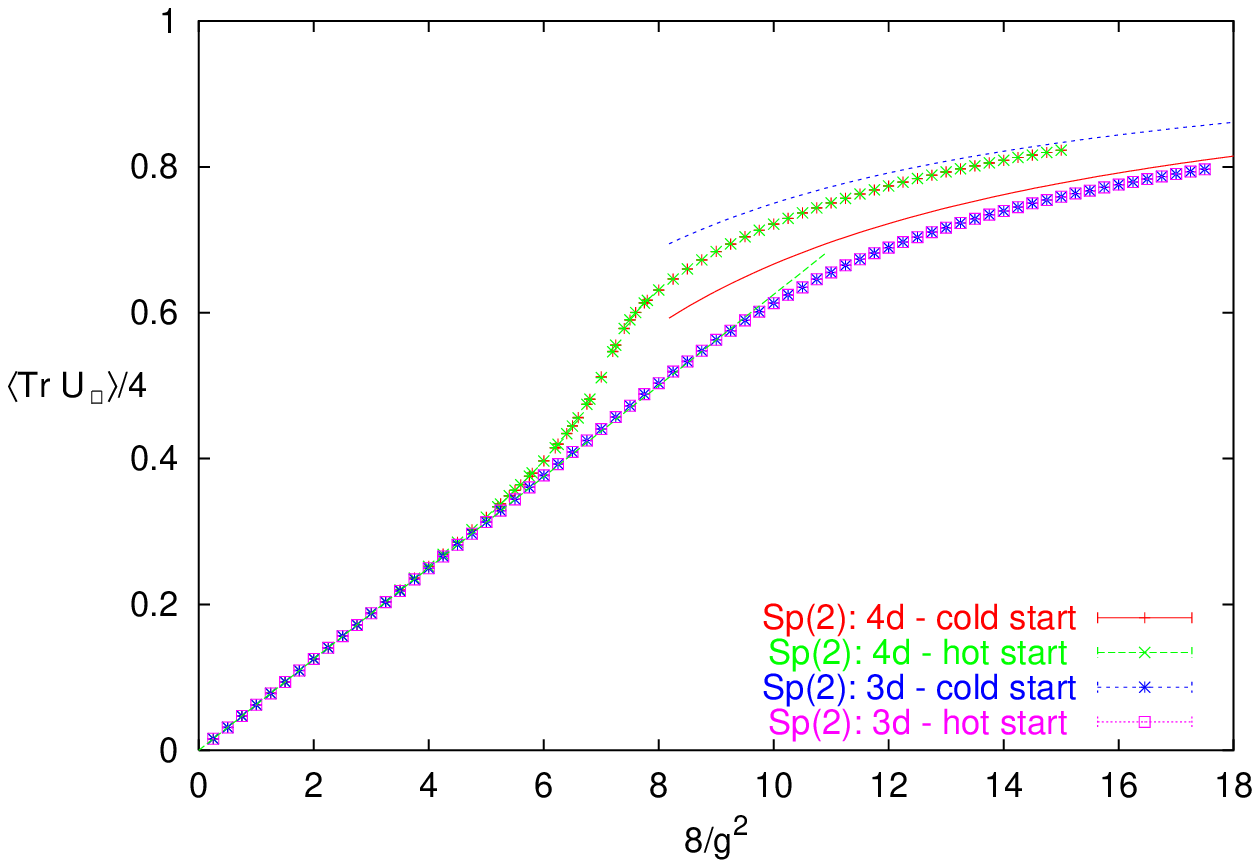,width=15cm}
\end{center}
\caption{\it Monte Carlo data from hot and cold starts for the plaquette in 
$(2+1)$-d and $(3+1)$-d $Sp(2)$ Yang-Mills theory compared to analytic results 
in the weak and strong coupling limits.}
\end{figure}

\begin{figure}[htb]
\begin{center}
\epsfig{figure=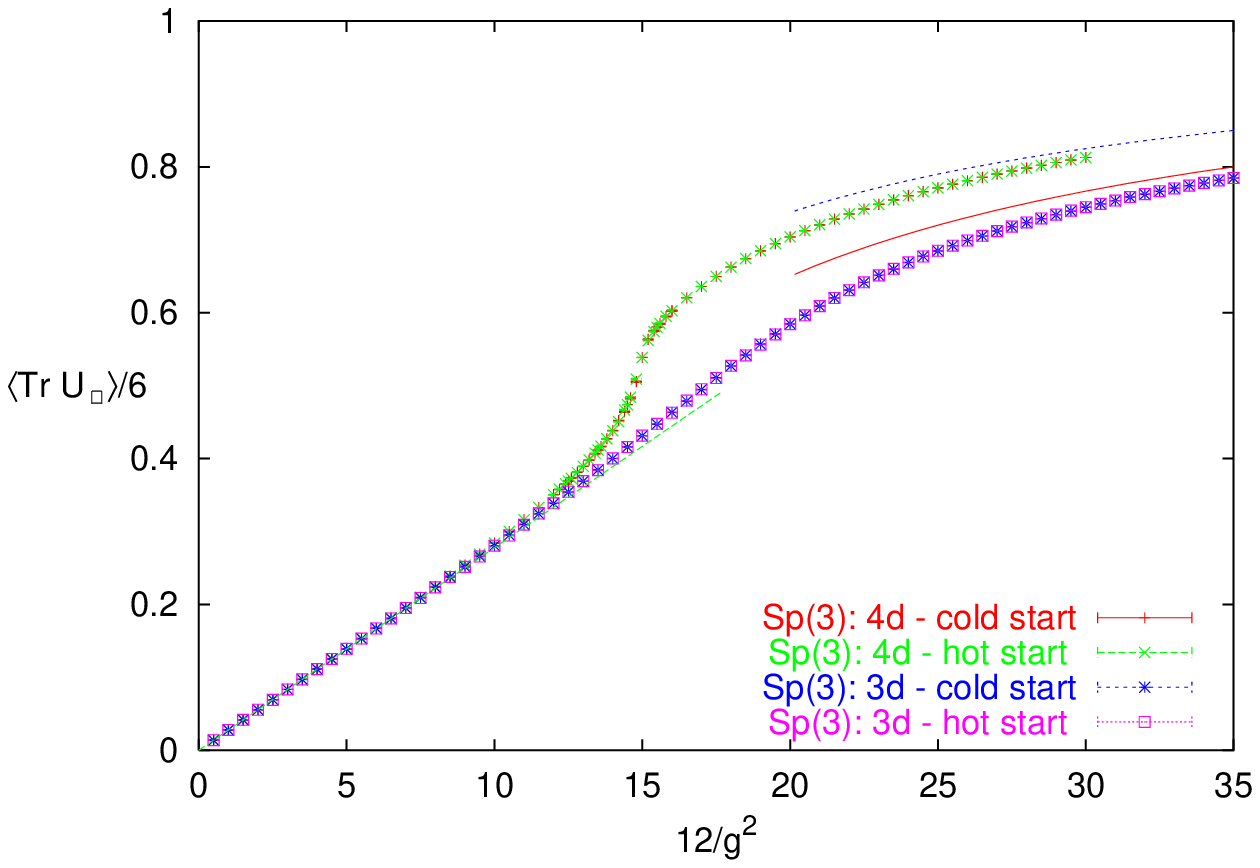,width=15cm}
\end{center}
\caption{\it Monte Carlo data from hot and cold starts for the plaquette in 
$(2+1)$-d and $(3+1)$-d $Sp(3)$ Yang-Mills theory compared to analytic results 
in the weak and strong coupling limits.}
\end{figure}

\begin{table}
\begin{center}
\begin{tabular}{|c|c|c||c|c|c|}\hline
\multicolumn{3}{|c||}{$Sp(2)$} & \multicolumn{3}{c|}{$Sp(3)$} \\ \hline\hline
$8/g^2$ &   3-d  &  4-d & $12/g^2$ &   3-d  &  4-d \\ \hline
\hline 
\,  1.0   &   0.0623(4)  &   0.06240(7) \,    &
\,  2.0   &   0.0554(3)  &   0.05555(5) \,    \\
\hline
\,  2.0   &   0.1251(4)  &   0.12510(9) \,   &
\,  4.0   &   0.1113(3)  &   0.11118(5) \,    \\
\hline
\,  3.0   &   0.1878(5)  &   0.18796(9) \,    &
\,  6.0   &   0.1670(3)  &   0.16722(5) \,    \\
\hline
\,  4.0   &   0.2495(5)  &   0.25213(9) \,    &
\,  8.0   &   0.2233(3)  &   0.22420(5) \,    \\
\hline
\,  5.0   &   0.3129(6)  &   0.31960(10)      &
  10.0  &   0.2805(4)  &   0.28413(5) \,    \\
\hline
\,  6.0   &   0.3770(6)  &   0.39670(20)      &
  12.0  &   0.3387(4)  &   0.35052(7) \,    \\
\hline
\,  7.0   &   0.4404(5)  &   0.5115(5) \,  \, &
  14.0   &  0.4001(4)  &   0.43825(30)      \\
\hline
\,  8.0   &   0.5034(5)  &   0.63110(20)  &
  16.0  &   0.4631(4)  &   0.60250(20)      \\
\hline
\,  9.0   &   0.5629(5)  &   0.68383(13)      &
  18.0   &  0.5270(4)  &   0.66252(7)  \,   \\
\hline
  10.0  &   0.6130(4)  &   0.72146(10)      &
  20.0   &  0.5843(4)  &   0.70390(7)  \,   \\
\hline
  11.0  &   0.6553(4)  &   0.75057(9) \,    &
  22.0  &   0.6310(3)  &   0.73540(6) \,    \\
\hline
  12.0  &   0.6892(4)  &   0.77389(7) \,    &
  24.0  &   0.6690(4)  &   0.76050(5) \,    \\
\hline
  13.0  &   0.7167(4)  &   0.79300(5) \,    &
  26.0  &   0.6990(3)  &   0.78105(6) \,    \\
\hline
  14.0  &   0.7396(3)  &   0.80913(5) \,    &
  28.0  &   0.7237(3)  &   0.79827(5) \,    \\
\hline
  15.0  &   0.7591(4)  &   0.82290(3) \,    &
  30.0  &   0.7446(2)  &   0.81290(4) \,    \\
\hline
\end{tabular}  
\caption{\it Plaquette expectation values 
$\langle \mbox{Tr}\,\, U_\Box \rangle/2N$ for 
$3$-d and $4$-d $Sp(2)$ and $Sp(3)$ 
  Yang-Mills theories on $8^3$ and $8^4$ lattices.}
\label{table1}
\end{center}
\end{table}

\subsection{Order of the Deconfinement Phase Transitions in $Sp(2)$ and $Sp(3)$
Yang-Mills Theories}

The $SU(2) = Sp(1)$ Yang-Mills theory has a second order deconfinement phase 
transition both in $2+1$ and in $3+1$ dimensions. Since all $Sp(N)$ groups 
have the same center $\Z(2)$, one might have expected all $Sp(N)$ Yang-Mills 
theories to have second order deconfinement phase transitions. However, it 
should be noted that Svetitsky and Yaffe made no statement about the order of 
the phase transition. Their conjecture just states that, if the transition is 
second order, it should be in the Ising universality class.

In order to investigate the order of the deconfinement phase transition, we 
have simulated $(2+1)$-d $Sp(2)$ Yang-Mills theory at finite temperature for 
$N_t = 2, 4$, and 6, at various spatial sizes ranging from $L = 10$ to
100. The probability distribution of the Polyakov loop, depicted in figure 6, indeed
shows the characteristic features of a second order phase transition. In 
particular, approaching the phase transition from the confined side, 
fluctuations broaden the distribution, which then evolves into a two-peak 
structure on the deconfined side. The finite-size scaling analysis presented in
the next subsection indeed confirms that the transition is second order and in 
the 2-d Ising universality class.

\begin{figure}[htb]
\begin{center}
\epsfig{figure=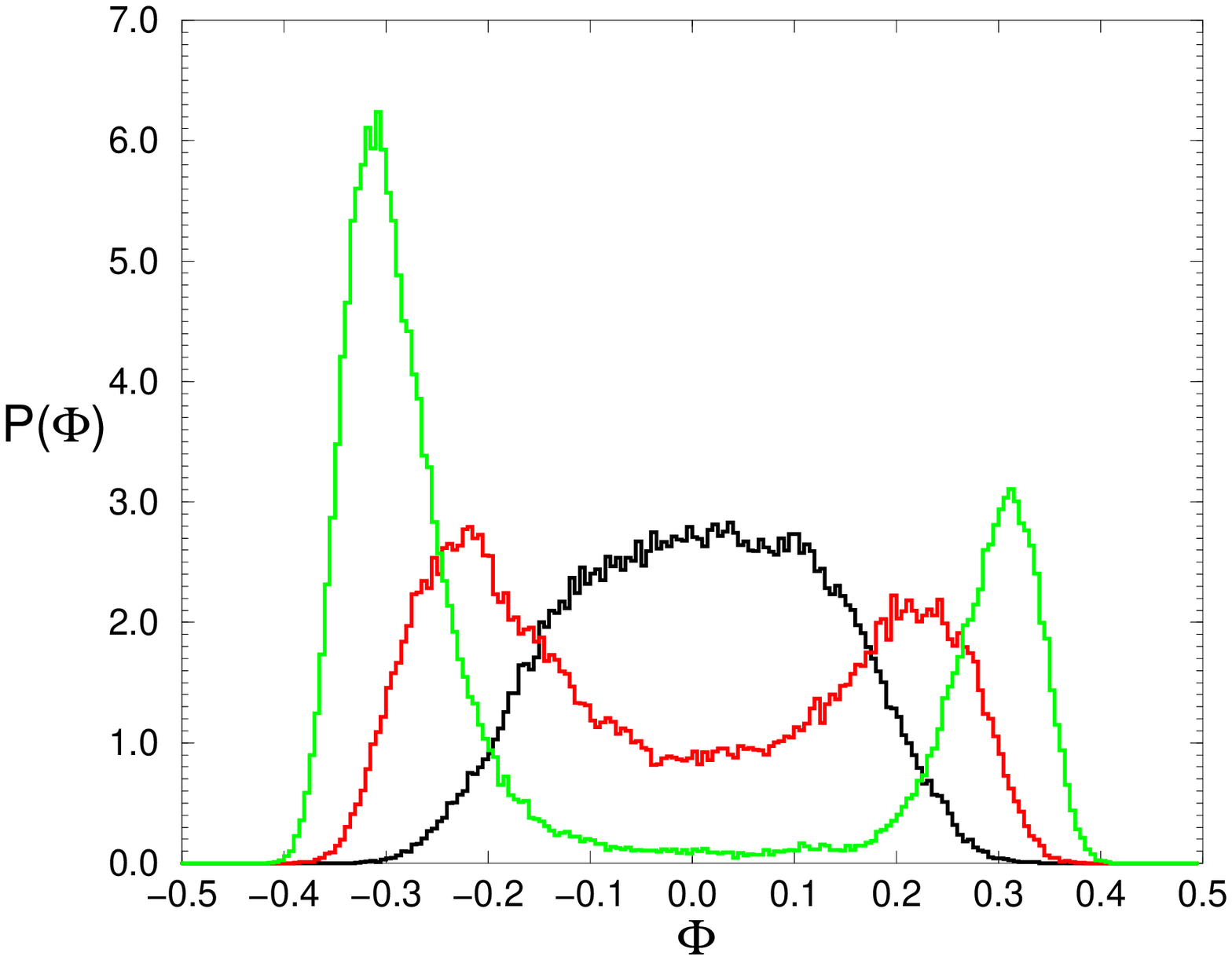,width=15cm}
\end{center}
\caption{\it Polyakov loop probability distributions for $(2+1)$-d $Sp(2)$
Yang-Mills theory on a $40^2 \times 2$ lattice at three different couplings
$8/g^2 = 10.3$, $10.4$, and $10.5$ close to the phase transition.}
\end{figure}

Interestingly, a corresponding study in $(3+1)$-d $Sp(2)$ Yang-Mills theory
shows a first order transition. In this case, we have performed numerical
simulations for $N_t = 2,3,4,5$, and 6 with $L = 8,10,...,20$. The probability 
distribution of the Polyakov loop, displayed in figure 7, clearly shows three 
maxima, indicating coexistence of the two deconfined and the confined phase. 
This signal becomes more pronounced on 
larger volumes. However, since tunneling events are then suppressed, due to 
limited statistics the deconfined peaks are sampled unevenly. 
\begin{figure}[htb]
\begin{center}
\epsfig{figure=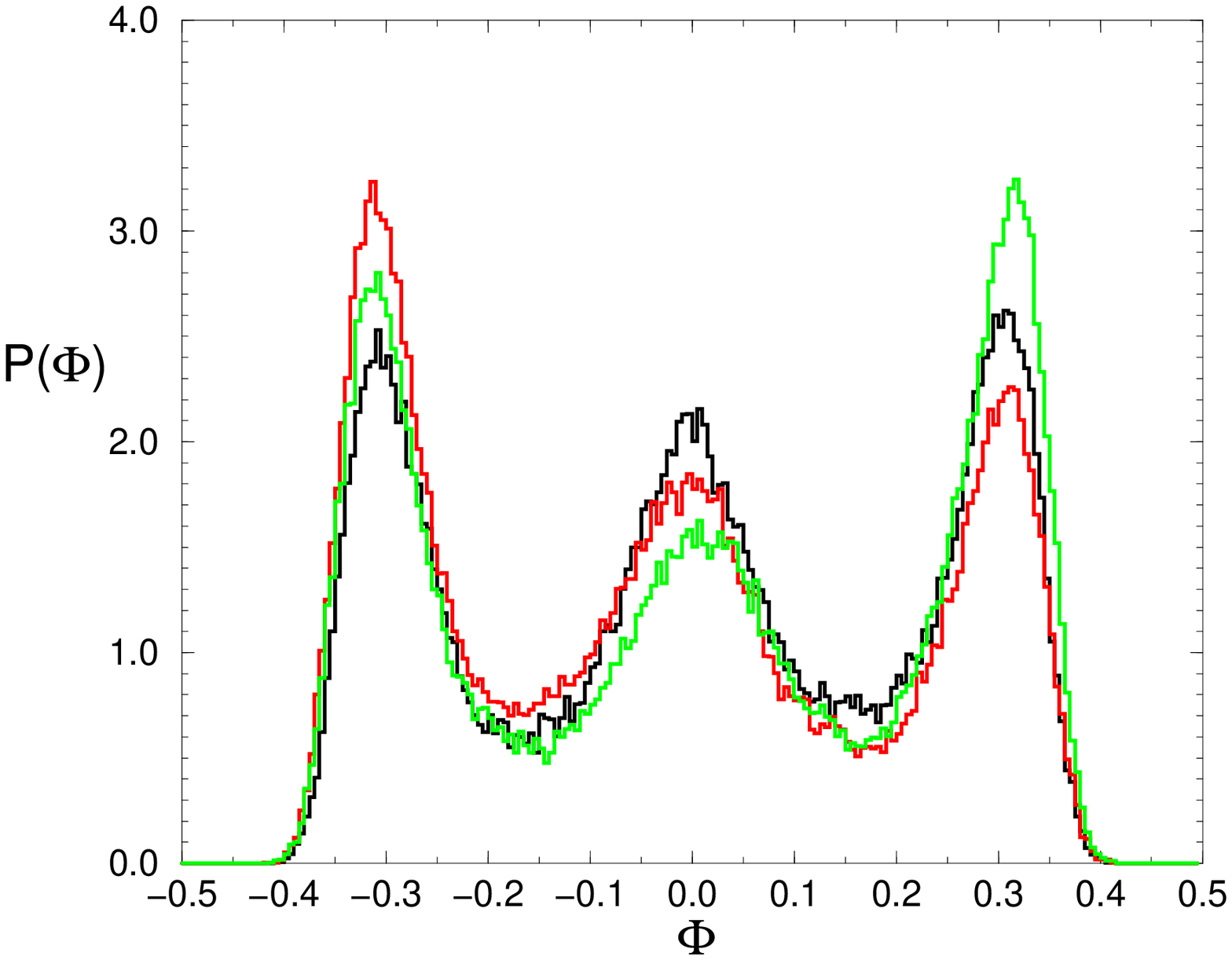,width=15cm}
\end{center}
\caption{\it Polyakov loop probability distributions for $(3+1)$-d $Sp(2)$
Yang-Mills theory on a $10^3 \times 2$ lattice at three different couplings
$8/g^2 = 6.464$, $6.465$, and $6.466$ close to the phase transition.}
\end{figure}
Figure 8 shows the Polyakov loop susceptibility $\chi$ as a function of $8/g^2$
for different spatial sizes $L$, keeping $N_t$ fixed. 
\begin{figure}[htb]
\begin{center}
\epsfig{figure=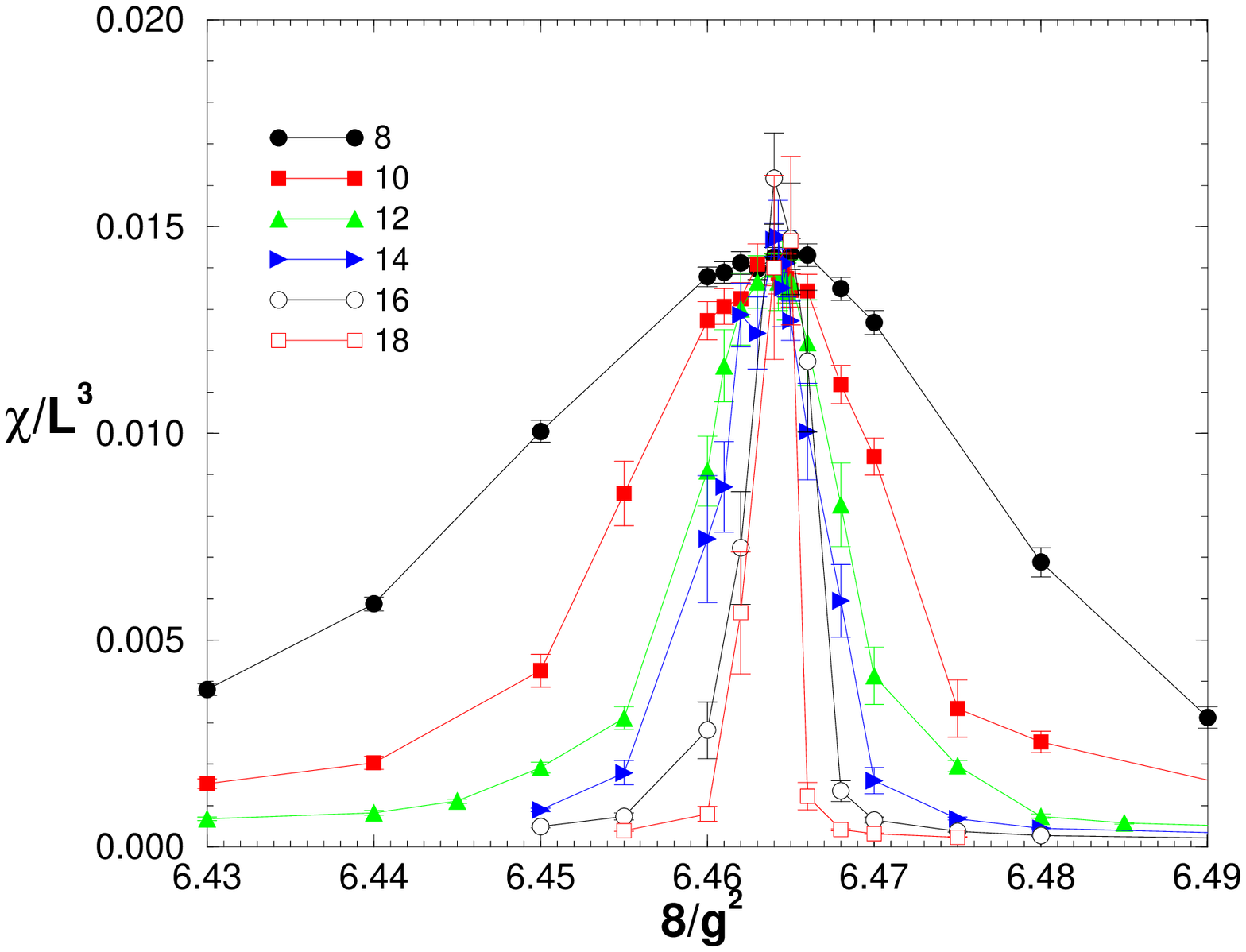,width=15cm}
\end{center}
\caption{\it Polyakov loop susceptibility per spatial volume, $\chi/L^3$, in
$(3+1)$-d $Sp(2)$ Yang-Mills theory for $L = 8,10,...,18$ and $N_t = 2$ as a 
function of the coupling $8/g^2$.}
\end{figure}
At $8/g_c^2 = 6.4643(3)$ the resulting behavior $\chi \sim L^3$ quantitatively
confirms the first order nature of the transition. Figure 9 shows the
maximum of the specific heat per volume, $C_V^{max}/L^3$, as a function of the inverse
volume $1/L^3$. The linear behavior is characteristic of a first order
phase transition. A linear extrapolation of $C_V^{max}/L^3$ to the infinite
volume limit (see eq.(\ref{speclat-heat})) is consistent with a direct
measurement of the latent heat $L_H$, which again supports the first
order nature of the transition. 
\begin{figure}[htb]
\begin{center}
\epsfig{figure=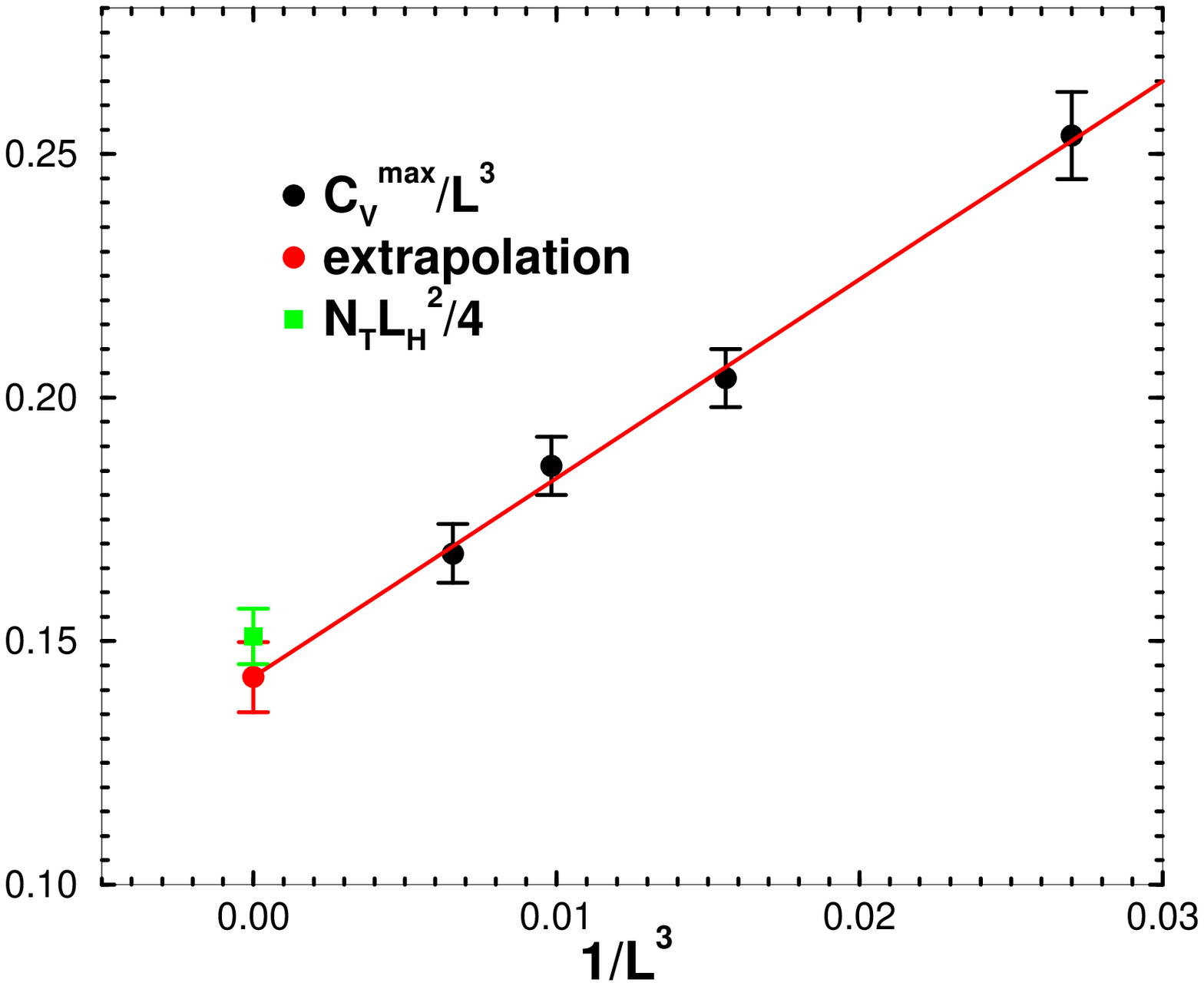,width=15cm}
\end{center}
\caption{\it The maximum of the specific heat per volume,
  $C_V^{max}/L^3$, as a function of the inverse 
volume $1/L^3$ for $(3+1)$-d $Sp(2)$ Yang-Mills theory. The linear
extrapo- lation to the infinite volume limit is in agreement with the
measured latent heat $L_H$.}
\end{figure}
Hence, despite the fact that the 3-d Ising universality class is
available, $(3+1)$-d $Sp(2)$ Yang-Mills theory does not fall into it
and instead displays non-universal first order behavior. As the
spatial dimension is increased from two to three, the coefficient $b$
of the $\Phi^4$ term in the effective potential $V(\Phi)$ of
eq.(\ref{effpot}) changes sign, thus driving the transition first
order. This effect does not happen in the Ising model or in $SU(2) =
Sp(1)$ Yang-Mills theory, but it does happen in $Sp(2)$ Yang-Mills
theory. One may wonder if this effect is a lattice artifact. As we
will see later, this seems not to be the case. Our data indicate that
the $(3+1)$-d $Sp(2)$ Yang-Mills theory has a first order
deconfinement phase transition even in the continuum limit.  

We have also investigated the deconfinement phase transition in $Sp(3)$ 
Yang-Mills theory, both in $2+1$ and in $3+1$ dimensions. Figure 10 shows 
the Polyakov loop distribution on a $40^2 \times 2$ lattice at $12/g^2 = 21.375$.
Three distinct peaks are clearly visible, indicating a first order phase
transition. Interestingly, compared to the $(2+1)$-d $Sp(2)$ case, the phase
transition has changed from second to first order. We attribute this to the
larger size of the group $Sp(3)$. The larger number of $Sp(3)$ gluons in the 
deconfined phase increases the difference between the relevant degrees of 
freedom on the two sides of the phase transition, thus driving the transition
first order. Similar behavior has been observed in $3+1$ dimensions. 
Figure 11 displays the Monte Carlo time history of the Polyakov loop on a 
$6^3 \times 2$ lattice at $12/g^2 = 13.83$ with multiple tunneling events between
the coexisting confined and deconfined phases. We have not attempted
to extrapolate our $Sp(3)$ results to the continuum limit. Since the
phase transition is very strongly first order at finite lattice
spacing, we expect that it remains first order in the continuum
limit. 

\begin{figure}[htb]
\begin{center}
\epsfig{figure=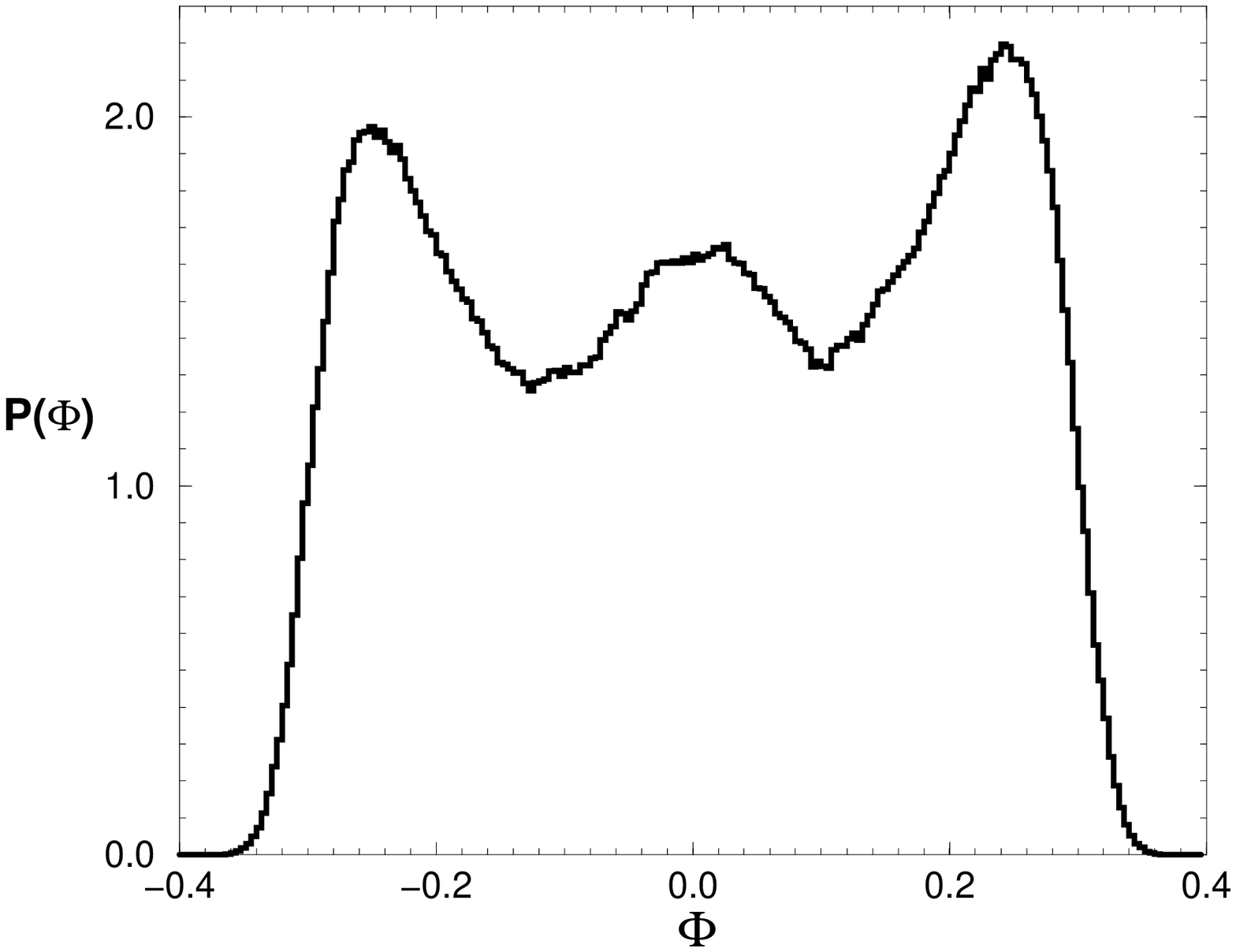,width=15cm}
\end{center}
\caption{\it Polyakov loop probability distributions for $(2+1)$-d $Sp(3)$
Yang-Mills theory on a $40^2 \times 2$ lattice at $12/g^2 = 21.375$.}
\end{figure}

\begin{figure}[htb]
\begin{center}
\epsfig{figure=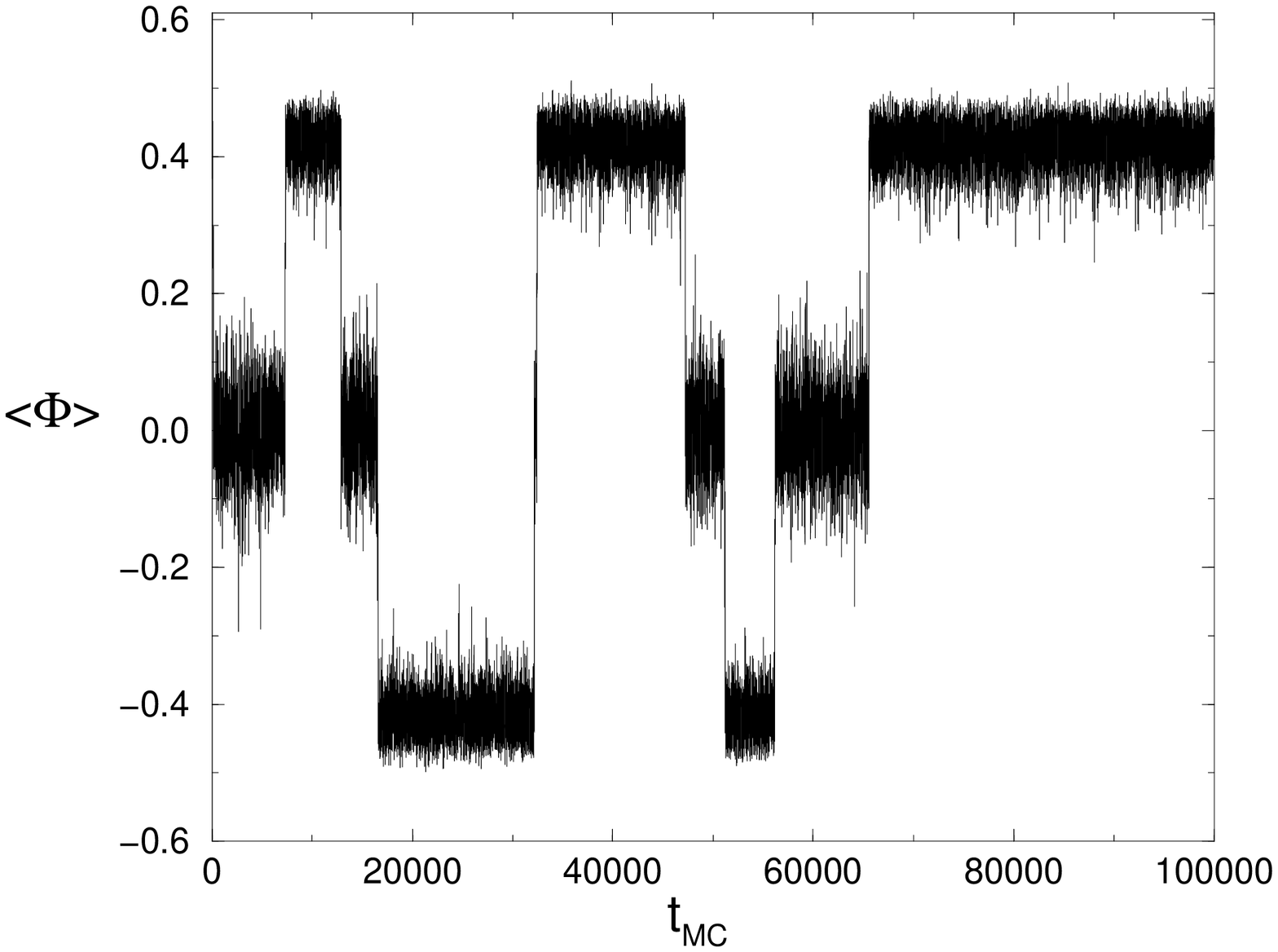,width=15cm}
\end{center}
\caption{\it Monte Carlo time history of the Polyakov loop in $(3+1)$-d $Sp(3)$
Yang-Mills theory on a $6^3 \times 2$ lattice at $12/g^2 = 13.83$.}
\end{figure}

\subsection{Finite-Size Scaling Analysis of the $(2+1)$-d $Sp(2)$ Deconfinement
Phase Transition}

A second order phase transition is characterized by a correlation length 
$\xi$ which diverges as the critical point is approached
\begin{equation}
\xi \sim x^{-\nu} = \left(\frac{g_c^2}{g^2} - 1\right)^{-\nu}.
\end{equation}
Here $x$ is a measure of the distance from criticality and the exponent
$\nu$ is particular to the universality class of the phase transition. Note 
that the critical coupling $g_c$ depends on $N_t$, which is kept fixed in the
finite-size scaling analysis. The universality class is determined by the
dimensionality of space and by the underlying symmetries. As well as the 
correlation length, the order parameter and the susceptibility also have 
particular power-like behavior close to the critical point
\begin{equation}
\langle |\Phi| \rangle \sim x^\beta, \ \chi \sim x^{-\gamma}.
\end{equation}
The exponents $\nu, \beta$, and $\gamma$ characterize the critical behavior and
are specific to a given universality class.

A finite system cannot have a phase transition in the sense of non-analytic 
behavior. In particular, the correlation length stays finite because it is 
limited by the system size. The correlation length sets a natural
distance scale and the ratio $L/\xi$ specifies the spatial size $L$ of
the system in these units. Close to criticality one can define the scaled variable
$y = x L^{1/\nu} \sim (L/\xi)^{1/\nu}$. Then, as the spatial volume increases, 
physical quantities approach criticality in a specific way. Finite-size scaling
is a method that relates the finite volume scaling behavior to the universal 
properties of the critical system in the thermodynamic limit. For instance, the
order parameter behaves as
\begin{equation}
\langle |\Phi| \rangle \sim L^{-\beta/\nu} F(x L^{1/\nu}),
\end{equation}
where $F(y)$ is a universal finite-size scaling function. By measuring the 
order parameter $\langle |\Phi| \rangle$
for a variety of spatial volumes and couplings $8/g^2$ close to the critical 
point, one can determine the critical exponents of the universality class. In 
order to check if our system is in the Ising universality class, we first 
assume the 2-d Ising critical exponents, $\nu = 1, \beta = 1/8$, and 
$\gamma = 7/4$, and only vary the critical coupling $8/g_c^2$. If corrections 
to scaling are small, then all the data should fall onto a universal curve for 
the correct value of $8/g_c^2$. Figure 12 is the finite-size scaling plot for 
$\langle |\Phi| \rangle L^{\beta/\nu}$ obtained with fixed temporal extent 
$N_t = 2$ and spatial volumes ranging from $L=26$ to 100. The data fall 
beautifully onto a universal curve for $8/g_c^2 = 10.45(1)$, indicating that
$(2+1)$-d $Sp(2)$ Yang-Mills theory has a second order deconfinement transition
in the universality class of the 2-d Ising model. Figure 12 also
contains data from the $SU(2)$ Yang-Mills theory, which is known to
fall in the 2-d Ising universality class. Indeed the data from the two
theories fall on the same universal curve.

\begin{figure}[htb]
\begin{center}
\epsfig{figure=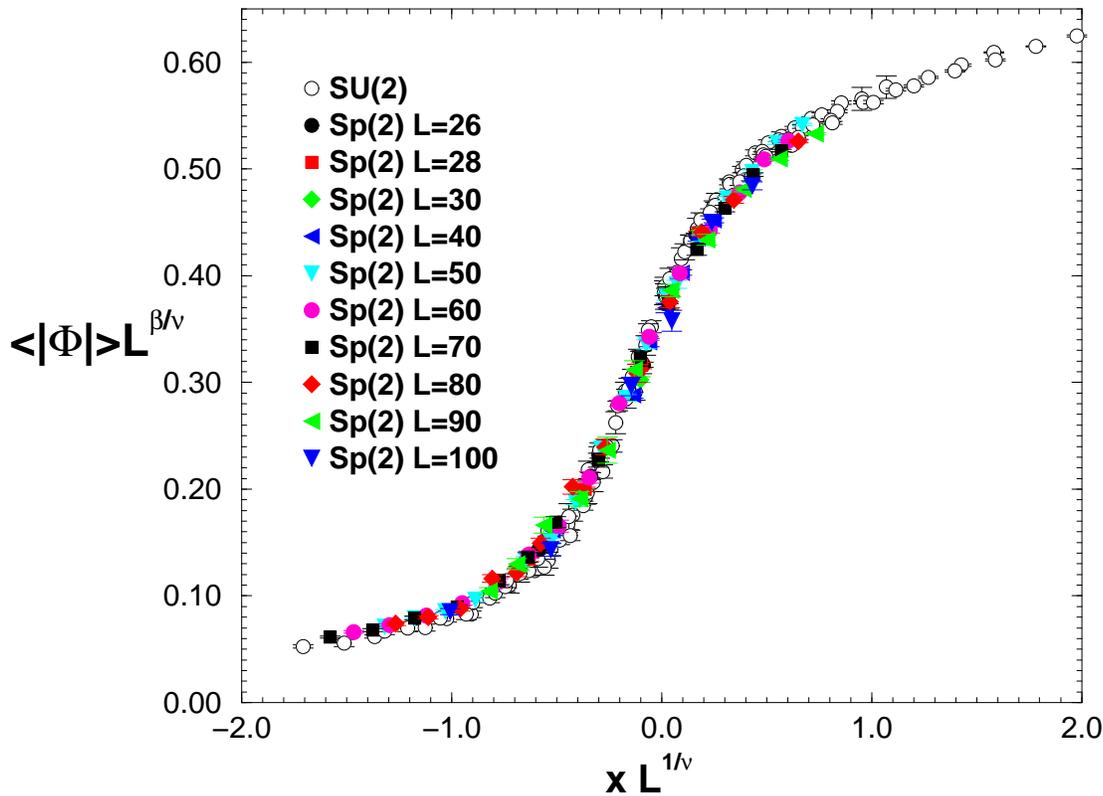,width=15cm}
\end{center}
\caption{\it Finite-size scaling plot for $\langle |\Phi| \rangle  L^{\beta/\nu}$ 
in $(2+1)$-d $Sp(2)$ Yang-Mills theory at $N_t = 2$. Some $SU(2)$ data
are also included.} 
\end{figure}

The value of the Binder cumulant $g_R$ at the critical point $x = 0$
is independent of the spatial size $L$, provided that corrections to
scaling are negligible (which is true for sufficiently large volumes). Just
like the critical exponents, the value $g_R(x=0)$ is another
characteristic of the universality class. For the 2-d Ising model, its value
is $g_R(x=0)=-1.837(8)$ \cite{Jan93}. From Figure 13, we see that this is
in good agreement with the value measured in $(2+1)$-d $Sp(2)$
Yang-Mills theory, further evidence that the phase transition belongs
to the expected universality class. As $g_R(x L^{1/\nu})$ changes
rapidly over a small range around the critical point, this can be used
to determine quite accurately the critical coupling $8/g_c^2$ as a function of
the temporal extent $N_t$ without having to assume the 
values of the universal exponents. At the critical coupling, the measured 
values of $g_R$ are independent of $L$ and in very good agreement with the 2-d 
Ising value quoted above. Figure 14 shows the finite-size scaling
function of the susceptibility $ \chi L^{-\gamma/\nu}$. Again all data
fall on a universal curve consistent with the critical exponents of the
2-d Ising model.

\begin{figure}[htb]
\begin{center}
\epsfig{figure=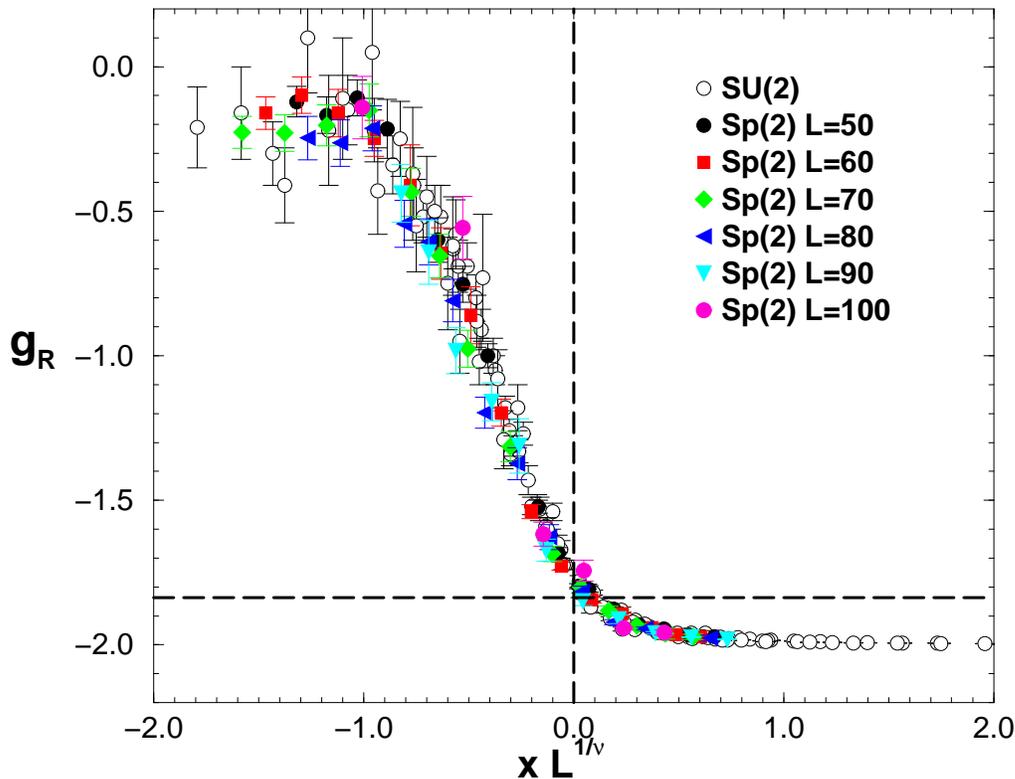,width=15cm}
\end{center}
\caption{\it Finite-size scaling plot for the Binder cumulant $g_R$ in 
$(2+1)$-d $Sp(2)$ Yang-Mills theory at $N_t = 2$. Some $SU(2)$ data are
also included. The intersection of the dashed lines indicates the
Ising value $g_R (x=0)=-1.837(8)$.}
\end{figure}

\begin{figure}[htb]
\begin{center}
\epsfig{figure=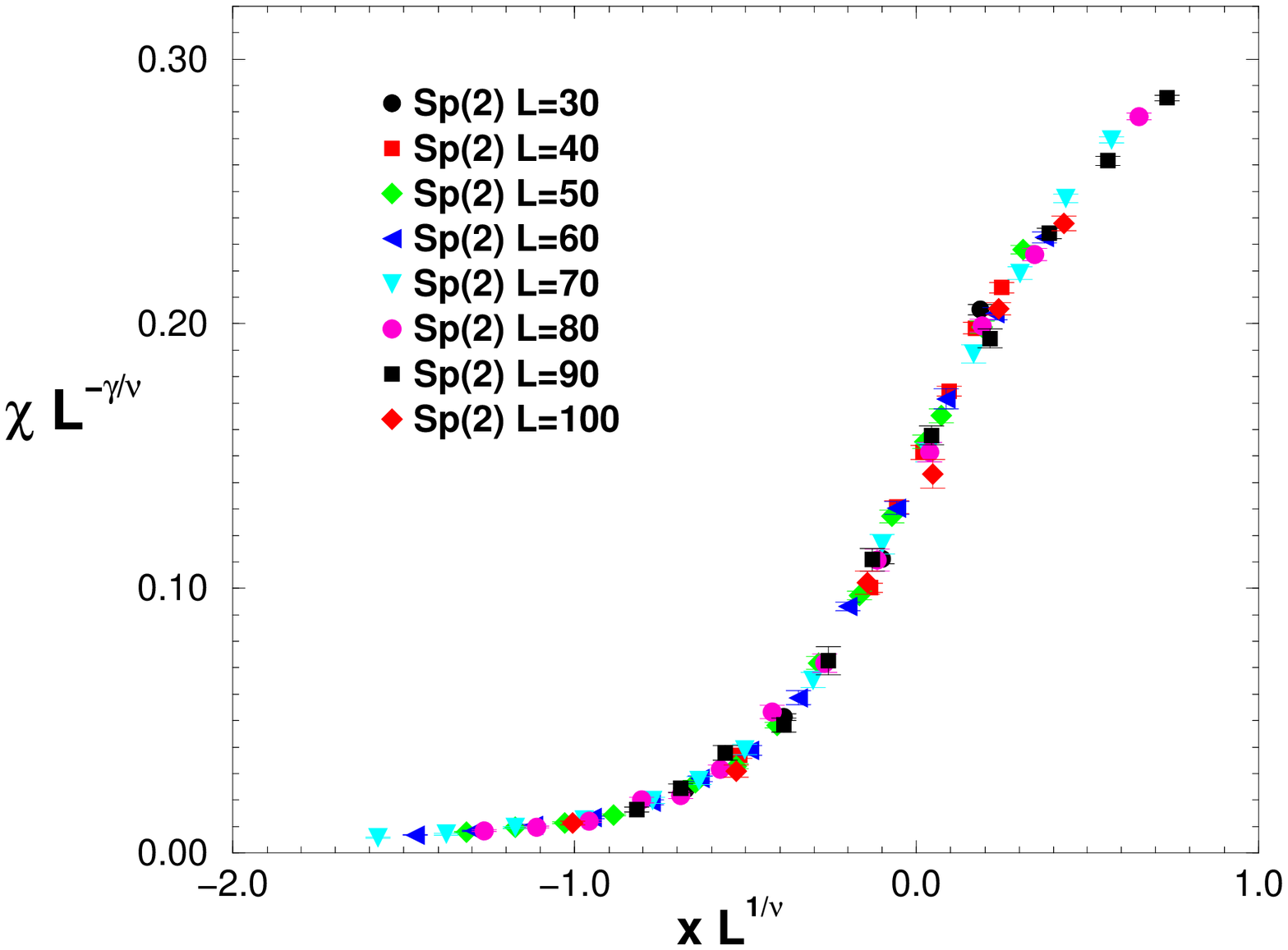,width=15cm}
\end{center}
\caption{\it Finite-size scaling plot for $\chi L^{-\gamma/\nu}$ in 
$(2+1)$-d $Sp(2)$ Yang-Mills theory at $N_t = 2$.}
\end{figure}

\subsection{Continuum Limit of $(3+1)$-d $Sp(2)$ Yang-Mills Theory}

We have presented numerical results indicating a first order deconfinement 
phase transition for $(3+1)$-d $Sp(2)$ and for $(2+1)$-d and $(3+1)$-d $Sp(3)$ 
Yang-Mills theory. In the $Sp(2)$ case the Euclidean time extents 
$N_t = 2,3,4,5$, and 6 have been investigated. First order signals were obtained 
in all cases and the critical couplings $8/g_c^2$ have been determined for
each of these $N_t$ values and are listed in table 2.

\begin{table}[htb]
\begin{center}
\begin{tabular}[h]{|c||c|c|c|c|c|}
\hline
$1/T_c$ & 2 & 3 & 4 & 5 & 6 \\
\hline
$8/g^2_c$ & 6.4643(3) & 7.1228(4) & 7.339(1) & 7.486(4) & 7.611(14) \\
\hline
\end{tabular}
\caption{\it Critical couplings $8/g^2_c$ for $N_t=2,3,4,5,6$ for
  $(3+1)$-d $Sp(2)$ Yang-Mills theory.}
\label{table2}
\end{center}
\end{table}

In this subsection we ask if the transition remains
first order in the continuum limit. To test for scaling, we have measured the 
string tension $\sigma$ in the zero temperature limit close to the critical 
couplings $8/g_c^2$. In the scaling regime the dimensionless ratio 
$T_c/\sqrt{\sigma}$ --- where $T_c = 1/N_t$ is the critical temperature in
lattice units --- should become independent of $N_t$. We have determined the 
string tension $\sigma$ (as well as the parameters $V_0$ and $c$ of 
eq.(\ref{pot})) from the static quark potential $V_{QQ}(\vec R)$ exploiting the
L\"uscher-Weisz multi-level algorithm \cite{Lue01}. An example of a typical Polyakov 
loop correlator together with a fit based on eq.(\ref{pot}) is depicted in figure 15. 
\begin{figure}[htb]
\begin{center}
\epsfig{figure=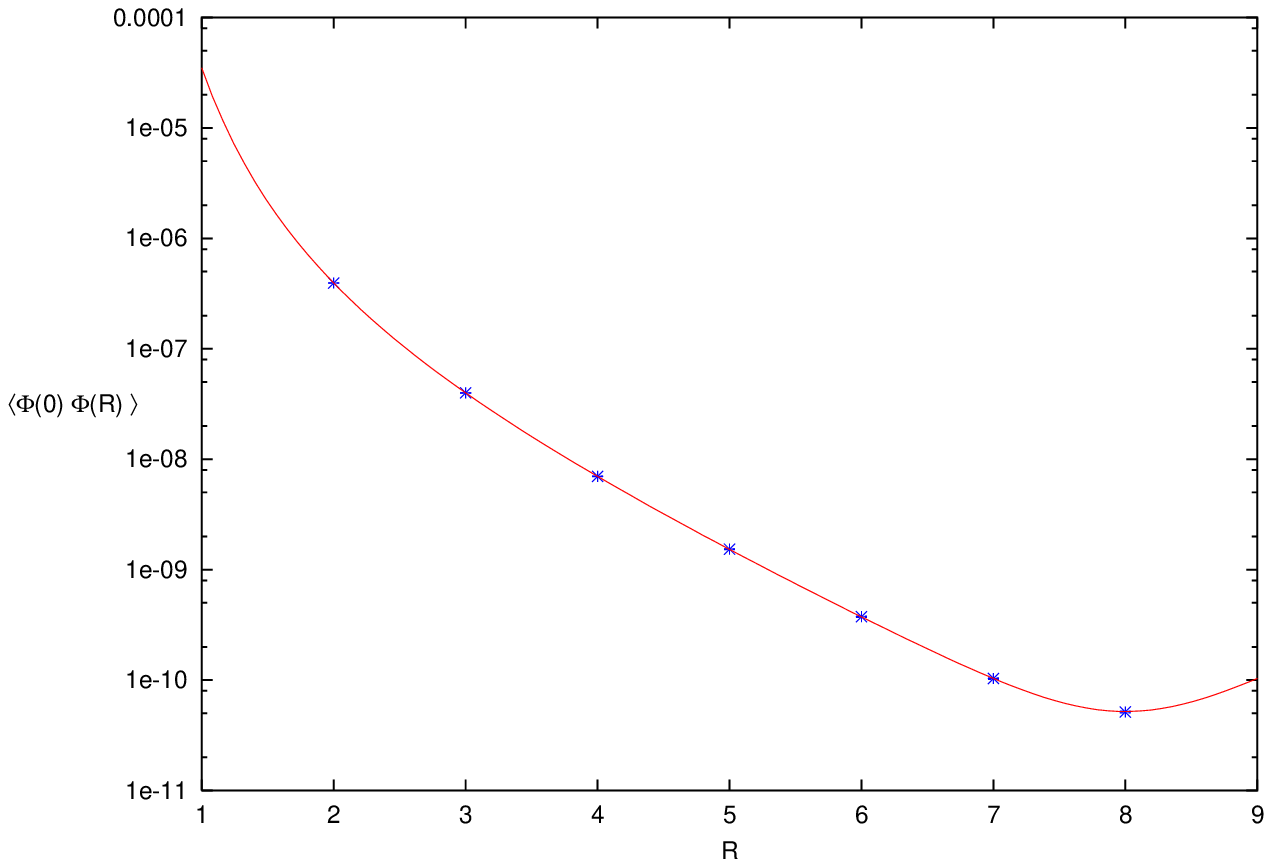,width=15cm}
\end{center}
\caption{\it Polyakov loop correlation function in $(3+1)$-d $Sp(2)$ Yang-Mills
theory at $8/g^2 = 7.635$ for $L = 16$ and $N_t = 20$. The line is a fit to 
$A [\exp(- N_t V_{QQ}(R)) + \exp(- N_t V_{QQ}(L-R))]$ with the static quark 
potential $V_{QQ}(R)$ given by eq.(\ref{pot}). The resulting string
tension  $\sigma = 0.0597(2)$ is quite accurately measured.}
\end{figure}
The relevant numerical data are summarized in table 3. For $N_t =
2,3,4,$ and 5 the critical
coupling $8/g_c^2$ has been determined quite accurately. In particular, the
statistical error of the string tension is comparable with the
systematic uncertainty resulting from the error in $8/g_c^2$. This is not 
the case for $N_t = 6$ where the error in $8/g_c^2$ leads to a larger
systematic uncertainty in the string tension evaluated at the critical coupling.
Not surprisingly, the result for $N_t=2$ is not in the scaling
regime. Figure 16 shows the extrapolation of $T_c/\sqrt{\sigma}$ for
$N_t =3,4,$ and 5 to the continuum limit. As expected, the  
cut-off effects are consistent with proportionality to the lattice spacing 
squared. Indeed, our data seem to be in the scaling region, which indicates 
that the deconfinement phase transition of $(3+1)$-d $Sp(2)$ Yang-Mills theory 
remains first order in the continuum limit. Extrapolating to that
limit, we find $T_c/\sqrt{\sigma} = 0.6875(18)$. This value
is very close to the known result 
$T_c/\sqrt{\sigma} = 0.7091(36)$ for $SU(2) = Sp(1)$ \cite{Tep02}, 
but significantly larger than the $SU(3)$ result 
$T_c/\sqrt{\sigma} = 0.6462(30)$ \cite{Tep02}. This may suggest that
$SU(2) = Sp(1)$ Yang-Mills theory is closer to the large $N$ limit of $Sp(N)$ than to
the one of $SU(N)$. It would be interesting to study this question by
investigating $T_c/\sqrt \sigma$ in $Sp(N)$ Yang-Mills theory with $N\geq 3$.

It should be pointed out that we observe scaling of dimensionless
ratios of physical quantities but no asymptotic scaling of individual 
quantities with the perturbative $\beta$-function of the bare coupling
constant. For example, a change of scale by a factor of 2 from
$N_t=3$ to $N_t=6$ requires a shift in the critical bare coupling $8/g_c^2$ by
$7.611(14)-7.1228(4)=0.488(15)$. On the other hand, asymptotic scaling
with the 1-loop $\beta$-function of $Sp(N)$ Yang-Mills theory
corresponds to a shift in $4N/g^2$ by
\begin{equation}
\frac{11 N (N+1)}{6 \pi^2} \ln 2 \approx 0.773 \,\,\,\, \mbox{for} \ N = 2.
\end{equation}
Similar violations of asymptotic scaling are familiar from numerical
simulations of $SU(N)$ Yang-Mills theory \cite{Fin92}. In particular, asymptotic
scaling is expected to set in only for bare couplings much
closer to the continuum limit.

\begin{figure}[htb]
\begin{center}
\epsfig{figure=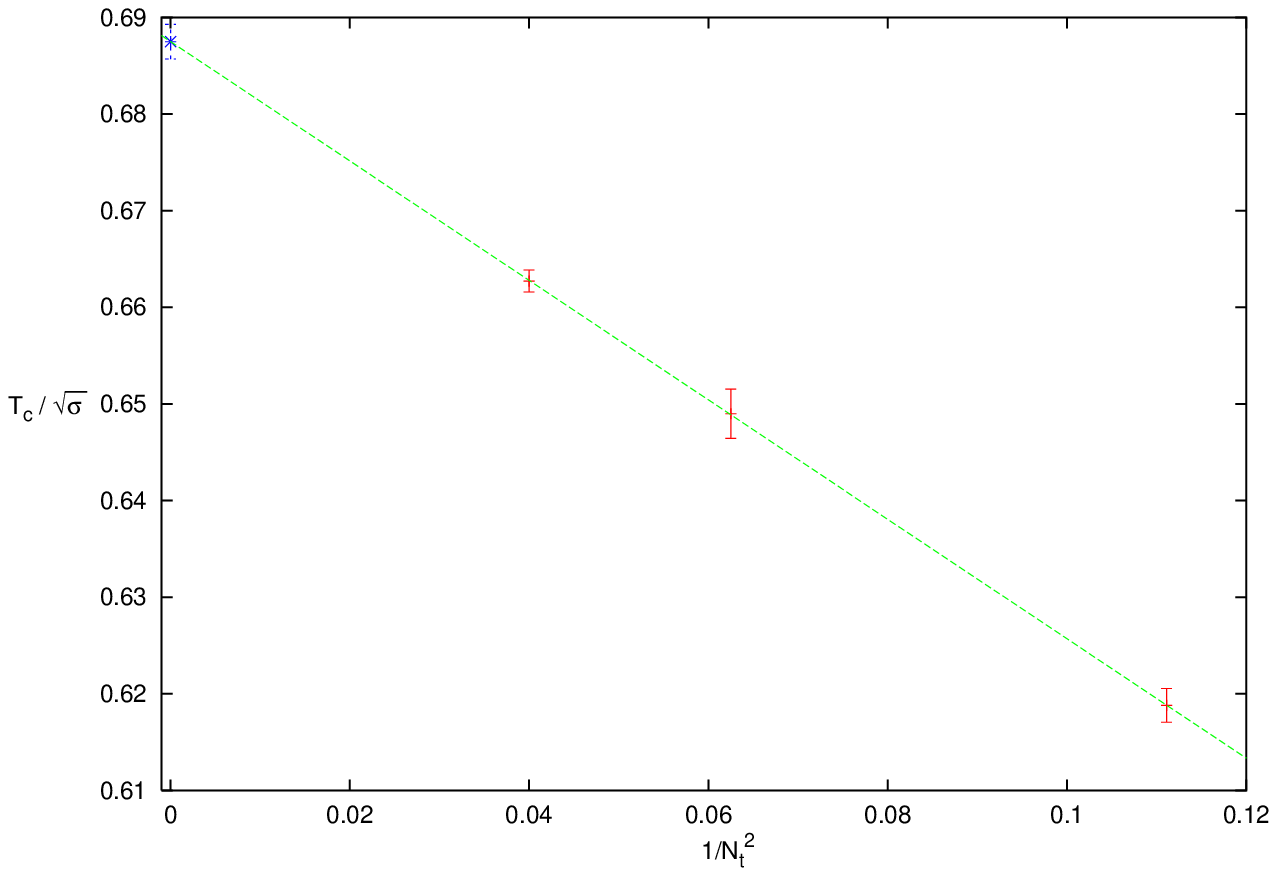,width=15cm}
\end{center}
\caption{\it Continuum extrapolation of the dimensionless ratio 
$T_c/\sqrt{\sigma}$.}
\end{figure}

\begin{table}[htb]
\begin{center}
\begin{tabular}[h]{|c|c|c|}
\hline
$8/g^2$ & $L^3\times N_t$ & $\sigma$ \\ 
\hline \hline
\, 6.4643 & $12^4$ & 0.71(1)\,\,\,\,\,\,\,\,\,  \\
\hline
7.123 & $12^4$ & 0.2902(16)     \\
\hline
7.340 & $12^4$ & 0.1484(12)     \\
\hline
7.490 & $12^4$ & 0.0904(11)     \\
      & $16^4$ & 0.0911(3) \,\  \\
\hline
7.620 & $16^4$ & 0.0617(3) \,\    \\
\hline
7.635 & $12^4$ & 0.0584(7)\,\,\, \\
      & $16^4$ & 0.0595(3)\,\,\, \\
      & $16^3 \times 20$ & 0.0597(2)\,\,\, \\
\hline
\end{tabular}
\caption{\it The string tension $\sigma$ for $(3+1)$-d $Sp(2)$ 
Yang-Mills theory at various couplings $8/g^2$.}
\label{table3}
\end{center}
\end{table}

\section{Conclusions}

It is interesting to systematically investigate the deconfinement phase
transition of Yang-Mills theories. The only case with rank 1 is $SU(2) = 
Spin(3) = Sp(1)$ Yang-Mills theory which has a second order deconfinement
phase transition with Ising critical exponents. The rank 2 groups include 
$SU(3)$, $SO(4) \simeq Spin(4) = SU(2) \otimes SU(2)$, $SO(5) \simeq Spin(5) = 
Sp(2)$, and $G(2)$. While $(3+1)$-d $SU(3)$ Yang-Mills theory has a first order 
deconfinement phase transition, in the $SU(2) \otimes SU(2)$ case the 
transition is second order and in the universality class of two decoupled 3-d
Ising models. The present study shows that in $(3+1)$-d $Sp(2)$ Yang-Mills 
theory the transition is again first order. This leaves $G(2)$ as the only 
unexplored case of rank 2. We have recently conjectured that, due to the 
triviality of its center, $G(2)$ Yang-Mills theory has no deconfinement phase 
transition --- just a crossover between a low- and a high-temperature regime 
\cite{Hol03}. It would be interesting to investigate this in numerical 
simulations. In particular, this would complete the numerical study of gauge 
groups of rank one and two. A systematic investigation could then proceed to 
the rank 3 groups $SO(6) \simeq Spin(6) = SU(4)$, $Sp(3)$, and $SO(7) \simeq 
Spin(7)$. As we now know, not only $SU(4)$ but also $Sp(3)$ Yang-Mills theory
has a first order deconfinement phase transition. Like $Sp(3)$, the group 
$Spin(7)$ also has the center $\Z(2)$. It would be interesting to see if 
$(3+1)$-d $Spin(7)$ Yang-Mills theory has a second order deconfinement phase 
transition in the 3-d Ising universality class. Since --- just like $Sp(3)$ ---
$Spin(7)$ is a large group (with 21 generators), we expect that it has a first 
order transition.

Let us summarize the results for the order of the deconfinement phase 
transition in $(3+1)$-d Yang-Mills theories. Based on lattice calculations with
the gauge groups $SU(3)$, $SU(4)$, $SU(6)$, and $SU(8)$, it seems natural to 
assume that all $SU(N)$ Yang-Mills theories with $N \geq 3$ have a first order 
transition. For $SU(2) = Sp(1) = Spin(3)$ the transition is second order. The 
same is true for $SU(2) \otimes SU(2) = Spin(4)$. Based on our $Sp(2)$ and 
$Sp(3)$ results, we expect that all $Sp(N)$ Yang-Mills theories with $N \geq 2$
have a first order deconfinement phase transition. Since for $SU(4) = Spin(6)$ 
the transition is again first order, we expect the same for all $Spin(N)$ with 
$N \geq 5$. For $E(6)$ with the center $\Z(3)$ the transition should be first 
order, for the same reason as for $SU(3)$: no universality class with $\Z(3)$ 
symmetry seems to exist in three dimensions. In addition, due to the large size
of $E(6)$, we would expect a first order transition even if a corresponding 
universality class was available. Again due to its large number of generators, 
and despite its center $\Z(2)$, we expect $E(7)$ Yang-Mills theory to have a 
first order transition. The remaining exceptional groups $G(2)$, $F(4)$, and 
$E(8)$ have a trivial center and thus need not have a deconfinement phase 
transition at all. For $G(2)$ we expect just a crossover, but we cannot rule 
out a first order phase transition. Consequently, only for 
$SU(2) = Sp(1) = Spin(3)$ and for its trivial extension 
$SU(2) \otimes SU(2) = Spin(4)$ the transition is second
order and in the 3-d Ising universality class. Svetitsky and Yaffe's universality 
arguments do not apply to the other Yang-Mills theories in $3+1$ dimensions. 

In $(2+1)$ dimensions a few Yang-Mills theories have a second order deconfinement 
phase transition: they include those with $Sp(2)$, $Spin(4)$, $SU(2)$
\cite{Tep93,Eng97}, $SU(3)$ \cite{Chr92,Eng97}, and, perhaps, $SU(4)$
\cite{Gro85,deF03b} gauge groups. It remains to be seen if other
$(2+1)$-d Yang-Mills theories with gauge group $Spin(N)$ with $N \geq
7$ or $SU(N)$ with $N \geq 5$ belong on this list. Due to the large
size of these groups we find this unlikely. For example, $Sp(3)$ and
$Spin(7)$ have 21 and $SU(5)$ has 24 generators. Since we find $(2+1)$-d $Sp(3)$ 
Yang-Mills theory to have a first order deconfinement phase transition, we 
expect the same for $Spin(N)$ with $N \geq 7$ and for $SU(N)$ with $N \geq 5$.

Our results suggest that the size of the gauge group  --- and not the center 
symmetry --- determines the order of the deconfinement phase transition. This 
should not be too surprising. The larger Lie groups have many generators and 
thus give rise to a large number of deconfined gluons. The number of confined 
glueball states, on the other hand, is essentially independent of the gauge 
group. For a large gauge group the drastic change in the number of relevant
degrees of freedom on the two sides of the deconfinement phase transition may 
easily drive it first order.

\section*{Acknowledgments}

We like to thank P.~de Forcrand, P.~Hasenfratz, O.~Jahn, P.~Minkowski,
F.~Niedermayer, and A.~Smilga for interesting discussions and
U.~Wenger for allowing us to use his reweighting code. This work is
supported by the Schwei\-ze\-ri\-scher Nationalfond, by the DOE under
the grant DOE-FG03-97ER40546 and by the European Community's Human
Potential Program under HPRN-CT-2000-00145 Hadrons/Lattice QCD, BBW
Nr.\ 99.0143.

\end{document}